\documentstyle[12pt,amssymb,epsf]{article}
\renewcommand{\theequation}{\arabic{section}.\arabic{equation}}

\def \thesection {\arabic{section}.}

\newcommand{\be}{\begin{equation}}
\newcommand{\ee}{\end{equation}}
\newcommand{\ba}{\begin{eqnarray}}
\newcommand{\ea}{\end{eqnarray}}
\newcommand{\baa}{\begin{eqnarray*}}
\newcommand{\btab}{\begin{tabular}}
\newcommand{\etab}{\end{tabular}}
\newcommand{\eaa}{\end{eqnarray*}}

\newcommand{\derleft}{\stackrel{\leftarrow}{D}}
\newcommand{\derright}{\stackrel{\rightarrow}{D}}

\def \labeltest #1 {\label{#1}}
\newcommand\re[1]{(\ref{#1})}

\newcommand\lr[1]{{\left({#1}\right)}}

\def \matrix #1 {\left(\begin{array}{cc} #1 \end{array}\right)}

\def \tr {\mbox{tr}}
\newcommand \vev [1] {\langle{#1}\rangle}

\newcommand \ket [1] {|{#1}\rangle}

\def \e {\mbox{e}}
\def \CO {{\cal O}}

\def\II{\hbox{{1}\kern-.25em\hbox{l}}}

\catcode`\@=11
\def\numberbysection{\@addtoreset{equation}{section}
                     \def\theequation{\thesection\arabic{equation}}}
\numberbysection

\begin{document}

\begin{titlepage}
\begin{flushright}
\begin{tabular}{l}
LPT--Orsay--99--68\\
SPbU--IP--99--12\\
hep-ph/9909539
\end{tabular}
\end{flushright}

\vskip1cm
\begin{center}
  {\large \bf
  Evolution equations for quark-gluon distributions in
  multi-color QCD and open spin chains \\}

\vspace{1cm}
{\sc S.\'{E}. Derkachov}${}^1$,
{\sc G.P.~Korchemsky}${}^2$
          and {\sc A.N.~Manashov}${}^3$
\\[0.5cm]

\vspace*{0.1cm} ${}^1$ {\it
Institut f\"ur Theoretische Physik,
Universit\"at Leipzig,\\
Augustusplatz 10, D-04109 Leipzig, Germany and
\\
Department of Mathematics,
St.-Petersburg Technology Institute,\\ St.-Petersburg, Russia
                       } \\[0.2cm]
\vspace*{0.1cm} ${}^2$ {\it
Laboratoire de Physique Th\'eorique%
\def\thefootnote{\fnsymbol{footnote}}%
\footnote{Unite Mixte de Recherche du CNRS (UMR 8627)},
Universit\'e de Paris XI, \\
91405 Orsay C\'edex, France
                       } \\[0.2cm]
\vspace*{0.1cm} ${}^3$ {\it
Department of Theoretical Physics,  Sankt-Petersburg State
University, \\ St.-Petersburg, Russia
                       }
\vskip0.8cm
{\bf Abstract:\\[10pt]} \parbox[t]{\textwidth}{
We study the scale dependence of the twist-3 quark-gluon parton
distributions using the observation that in the multi-color limit
the corresponding QCD evolution equations possess an additional
integral of motion and turn out to be effectively equivalent to
the Schr\"odinger equation for integrable open Heisenberg spin chain
model. We identify the integral of motion of the spin chain as a
new quantum number that separates different components of the
twist-3 parton distributions. Each component evolves independently
and its scale dependence is governed by anomalous dimension given
by the energy of the spin magnet. To find the spectrum of the QCD
induced open Heisenberg spin magnet we develop the Bethe
Ansatz technique based on the Baxter equation. The solutions to the
Baxter equation are constructed using different asymptotic methods
and their properties are studied in detail. We demonstrate that the
obtained solutions provide a good qualitative description of the spectrum
of the anomalous dimensions and reveal a number of interesting
properties. We show that the few lowest anomalous dimensions are
separated from the rest of the spectrum by a finite mass gap and
estimate its value.}

\vskip1cm

\end{center}

\end{titlepage}

\newpage

{\small \tableofcontents}

\newpage

\section{Introduction}

The evolution equations play an important r\^ole in the QCD
studies of hard processes as they allow to find the dependence of
hadronic observables on the underlying high-energy scales
\cite{report}. It has been recognized recently that in different
kinematical limits the QCD evolution equations possess an
additional hidden symmetry. Namely, the Regge asymptotics of
hadronic scattering amplitudes~\cite{Lip,FK} and the scale
dependence of the leading twist light-cone baryonic distribution
amplitudes \cite{BDM} reveal remarkable properties of
integrability. Both problems turn out to be intrinsically
equivalent to the Heisenberg closed spin magnet which is known to
be integrable $(1+1)-$dimensional quantum mechanical system and
their solutions can be found by applying powerful methods of
Integrable Models \cite{book}. In the present paper we continue
the study of integrable QCD evolution equations initiated in
\cite{BDM,BDKM} by considering the evolution equations for
twist-three quark-gluon distribution functions. These functions
describe the correlations between quarks and gluons in hadrons and
provide an important information about the structure of hadronic
states in QCD. They determine the twist-3 nucleon parton
distributions \cite{CFP} and the high Fock components of the meson
wave functions \cite{BBKT1} which in turn can be accessed
experimentally through the measurement of different asymmetries.

The twist-3 quark-gluon distribution functions $D(x_1,x_2,x_3;\mu)$
can be defined in QCD in terms of hadronic matrix elements of nonlocal gauge
invariant light-cone operators ${\cal F}(z_1,z_2,z_3)$ as
\cite{JJ}%
\footnote{The twist-3 contribution to the wave function is given
by a similar expression with $x_1+x_2+x_3=1$ and the matrix element
calculated between the vacuum and the meson state.}
\be
\vev {p,s |{\cal F}(z_1,z_2,z_3)|p,s }=\int_{-1}^1 {\cal D} x\
\e^{i(x_1 z_1+x_2z_2+x_3z_3)(pn)}
{D}(x_1,x_2,x_3;\mu)\,,
\label{dis-fun}
\ee
where ${\cal D}x=dx_1dx_2dx_3\delta(x_1+x_2+x_3)$ and $|p,s\rangle$ is the
nucleon state with momentum $p$ and spin $s$.
The distribution function ${D}(x_1,x_2,x_3;\mu)$ vanishes outside the region
$-1 < x_{1,2,3} <1$ and the scaling variables $x_1$, $x_2$ and $-x_3$
have a meaning of hadron momentum fractions carried by quark, gluon and
antiquark, respectively, in the infinite momentum frame \cite{JJ}.
For $0<x_i<1$ (or $-1<x_i<0$) the corresponding
parton belongs to the initial (or final) state nucleon.
The matrix element entering the l.h.s.\ of \re{dis-fun}
contains ultraviolet divergences that are subtracted at the scale $\mu$.
The normalization scale $\mu$ defines the transverse separation of the
partons and is determined by the hard scale of the underlying process
\cite{report}.

The operator ${\cal F}(z_1,z_2,z_3)$ describes the interacting system of
quark, antiquark and gluon ``living'' on the light-front $y_\mu=z n_\mu$
with $n^2=0$ and separated along the $z-$direction.
The twist-3 operator ${\cal F}(z_1,z_2,z_3)$ is given
by one of the following expressions \cite{SV,BB}
\ba
S^\pm_\perp(z_1,z_2,z_3) &=& \bar q(z_1n) \left[
\widetilde G_{\perp\nu}(z_2n) \pm
iG_{\perp\nu}(z_2n) \gamma_5 \right] n^\nu {\not\! n}  \, q(z_3n)
\label{S-def}
\\
T_\Gamma(z_1,z_2,z_3) &=&\bar
q(z_1n)n_\mu\sigma^{\mu\rho} n^\nu G_{\nu\rho}(z_2n)\Gamma\, q(z_3n)
\label{T-def}
\ea
with $\Gamma=\{\II\,, i\gamma_5\}$, $\widetilde G_{\mu\nu}=
\epsilon_{\mu\nu\rho\lambda} G^{\rho\lambda}/2$ being dual gluon
field strength and ``$\perp$'' denoting the ``transverse'' Lorentz
components orthogonal to the plane defined by the vectors $p_\mu$ and
$n_\mu$. The definition of the chiral-even and chiral-odd quark-gluons
distributions corresponding to the operators \re{S-def} and \re{T-def},
respectively, and their
relation to the twist-3 nucleon parton distributions are given in the
Appendix A. In order to avoid additional complication due to mixing
of the chiral-even operators $S^\pm$ with pure gluonic
operators we will assume quarks to be of different (massless) flavor.
It is implied that the gauge invariance of the operators \re{S-def} and
\re{T-def} is restored by including non-Abelian phase factors
$P\e^{i\int dx^\mu A_\mu(x)}$ connecting gluon strength field with
quark and antiquark fields.

In this paper we shall study the $\mu-$dependence of the distribution
functions \re{dis-fun} in the multi-color QCD. The standard way
of finding this dependence \cite{SV,BKL,KT} consists in expanding
nonlocal operator ${\cal F}(z_1,z_2,z_3)$ entering \re{dis-fun} over
the set of local gauge invariant composite operators
$
{\cal O}_{N,n}= \bar q (n\derleft)^n G(n\derright)^{N-n} q
$
with $n=0$, $...$, $N+1$ and $D_\mu=\partial_\mu-ig A_\mu$ being
the covariant derivative. In this way, one finds the evolution of the
moments
$
\int_{-1}^1 dx_1 dx_2 dx_3 \delta(x_1+x_2+x_3)\, x_1^{N-n}x_3^n\, {D}(x_i)
$
by diagonalizing a nontrivial $(N+1)\times (N+1)$ mixing matrix of the
operators ${\cal O}_{N,n}$. It becomes straightforward to calculate
numerically the spectrum of the anomalous dimensions for any given
$N$ \cite{KT} but the general analytical structure of the spectrum remains
unknown. To overcome this problem we apply the approach developed in
\cite{BDKM}. It allows to construct the basis of local twist-3
quark-gluon operators which evolve independently and whose
anomalous dimensions can be calculated analytically.

According to the definition, Eq.~\re{dis-fun}, the $\mu-$dependence of
the twist-3 quark-gluon distributions follows from the renormalization
group evolution of the corresponding nonlocal operators \re{S-def} and
\re{T-def}. To leading logarithmic accuracy one finds \cite{BB}
\be
\frac{d}{d\ln\mu}
{\cal F}(z_1,z_2,z_3) = -\frac{\alpha_s N_c}{2\pi}
{\cal H}_{\cal F} \cdot {\cal F}(z_1,z_2,z_3)
\label{RG}
\ee
with ${\cal H}_{\cal F}$ being some integral operator
describing a pair-wise interaction between quark, gluon and antiquark
with the light-cone coordinates $z_1$, $z_2$ and $z_3$, respectively.
The approach of \cite{BDKM} is based on the identification of the evolution
equation \re{RG} as one-dimensional Schr\"odinger equation for three
particles with a pair-wise interaction. The light-cone coordinates of the
quarks and gluon $y_i^\mu=z_i n^\mu$ define the spacial coordinates, $z_i$,
whereas the evolution parameter $\ln\mu^2$ plays the r\^ole of the evolution
time. The integration of three-body Schr\"odinger equation
\re{RG} becomes problematic unless there exists
an additional symmetry. It turns out that this is the case for the twist-3
quark-gluon correlation functions ${D}(x_i;\mu)$.
The corresponding  evolution equation \re{RG} possesses an additional
``hidden'' integral of motion in the multi-color limit and, as a consequence, it is
completely integrable \cite{BDM}.
As we will show, the underlying integrable
structure is identical to that for the integrable open {\it noncompact\/}
$SL(2;\mathbb R)$ spin chain models. Similar conclusions have been reached in
the recent publications \cite{B}. Our results partially overlap with
the results of \cite{BDM,B} and we shall comment on their relation below.

``Noncompactness'' of the spin group is a novel feature that QCD
brings into the theory of integrable open spin magnets and that
makes the analysis of the evolution equations interesting on its
own. We would like to notice that an additional motivation for
studying noncompact integrable open spin chains comes from the
recent analysis of the Regge asymptotics of quark-gluon scattering
amplitudes \cite{KK}. We also expect that these models (in their
generalized version) will inevitably appear in the analysis of the
evolution equations for higher twist quark-gluon correlation functions
\cite{CFP}.

In the present paper we develop an approach for solving noncompact
open spin chains models based on the Baxter equation and the fusion
relations and apply it to find the spectrum of the anomalous dimensions
of the twist-3 quark-gluon correlation functions in the leading
logarithmic approximation in the multi-color QCD.
As we will demonstrate this spectrum exhibits a number of interesting
properties. The few lowest anomalous dimensions turn out to be
separated from the rest of the spectrum by a finite ``mass gap'' and
can be calculated exactly. The remaining part of the spectrum can
be calculated using different asymptotic methods and
the obtained approximate expression agree well with the exact results
for large spin $N$.

The presentation is organized as follows. In Section 2 we explore the
conformal symmetry of the evolution equations to write the QCD evolution
kernels in the ``normal'' form in which their conformal invariance
becomes manifest. In Section 3 we construct a general
integrable open spin magnet and identify the values of parameters for
which the Hamiltonian of the model coincides with the QCD evolution
kernels for nonlocal operators $S^\pm_\perp$ and $T$. Section 4 is devoted to
diagonalization of the conserved charge by means of the Algebraic Bethe
Ansatz. Section 5 contains a derivation of the expression for the energy
of the open spin magnet, or equivalently the spectrum of the anomalous
dimensions of the quark-gluon correlation functions, in terms of the
solutions to the Baxter equation. In Section 6 we study the properties
of the Baxter equation and find its exact solutions. In Section 7 the
asymptotic methods are applied to describe a fine structure of the
spectrum. Summary of the main results is given in Section 8.
In Appendix A the relation between the quark-gluon distributions
and twist-3 nucleon structure functions is discussed.
Appendix B contains definition and some properties of the Wilson
polynomials.

\section{QCD evolution kernels}

In the multi-color limit the QCD evolution kernel ${\cal H}_{\cal F}$
entering \re{RG} gets contribution only from planar diagrams and
to one-loop level it is equal to the sum of the quark-gluon and
antiquark-gluon interaction kernels while the interaction between
quark and antiquark is suppressed by a color factor $1/N_c^2$.
The kernel ${\cal H}_{\cal F}$ depends on the choice of
nonlocal operator and for twist-3 nonlocal operators defined in
\re{S-def} and \re{T-def} it is given in the multi-color
limit by the following expressions \cite{BDM}
\ba
&&{\cal H}_T=\psi\left(J_{qg}+\frac32\right)+\psi\left(J_{qg}-\frac32\right)
          +\psi\left(J_{g\bar q}+\frac32\right)+\psi\left(J_{g\bar
          q}-\frac32\right)+\delta
\label{ker-T}
\\
&&{\cal H}_{S^+}=\psi\left(J_{qg}+\frac32\right)+\psi\left(J_{qg}-\frac32\right)
          +\psi\left(J_{g\bar q}+\frac12\right)+\psi\left(J_{g\bar
          q}-\frac12\right)+\delta
\label{ker-S}
\ea
with $\psi(x)=d\ln\Gamma(x)/dx$,
$\delta=-3/2+4\gamma_{_{\rm E}}$ and $J_{ik}$ being the operators
acting on the light-cone coordinates of the partons $(i,k=q,g,\bar q)$
and defined below in \re{L->J}. Here, the kernel ${\cal H}_T$
governs the RG evolution of the operators $T_\Gamma(z_i)$.
To one-loop level it does not depend on the choice of $\Gamma$ and
we will not indicate the subscript $\Gamma$ explicitly.
The kernel ${\cal H}_{S^-}$ is obtained from ${\cal H}_{S^+}$
by interchanging the coordinates of quark and antiquark and therefore
it will not be considered separately.

It is well known \cite{Conf} that to one-loop level the QCD
evolution kernels \re{ker-T} and \re{ker-S} inherit the conformal
symmetry of the bare QCD Lagrangian which is reduced on the
light-cone to its $SL(2,\mathbb{R})$ subgroup generating a
projective transformation $z\to (az+b)/(cz+d)$ on the line
$y_\mu=zn_\mu$ with $ad-bc=1$. Denoting the $SL(2,\mathbb{R})$
generators as $L_+$, $L_-$ and $L_0$ we define their action on the
quark and gluon fields $\Phi(z)\equiv \left\{\bar q(nz),\,
q(nz),\, n^\mu G_{\mu\nu}(nz)\right\}$ on the light-cone
$n^2_\mu=0$ as
\ba
L_{-,{k}} \Phi(z_{k}) &=& -\partial_{z_{k}}
\Phi(z_{k}) \nonumber
\\
L_{+,{k}} \Phi(z_{k}) &=&  \left(z^2_{k}\partial_{z_{k}}+
2 j_{k} z_{k}\right) \Phi(z_{k})
\label{L's}
\\
L_{0,{k}} \Phi(z_{k}) &=& \left(z_{k}\partial_{z_{k}}+j_{k}\right)
\Phi(z_{k})\,,
\nonumber
\ea
where $j_{k}=(l_{k}+s_{k})/2$ is the conformal spin of the field
$\Phi_{k}$ (${k}=\bar q, q, g$) defined as the sum of
its canonical dimension, $l_{k}$, and the projection of its spin on the
line $zn^\mu$, $\Sigma_{+-}\Phi_{k}=is_{k} \Phi_{k}$. In particular, using
$l_{\bar q}=l_q=3/2$, $l_g=2$ together with $s_{\bar q}=s_q=1/2$, $s_g=1$
we get the conformal spins as
\be
j_{\bar q}=j_q=1\,,\qquad j_g=3/2
\label{spins}
\ee
for (anti-)quark and gluon fields, respectively.

The operators $J_{qg}$ and $J_{\bar q g}$ entering \re{ker-S} and \re{ker-T}
are defined in terms of two-particle Casimir operators as follows
\be
L_{ik}^2=J_{ik}(J_{ik}-1)\,,\qquad
L_{ik,\alpha}=L_{i,\alpha}+L_{k,\alpha}
\label{L->J}
\ee
with $i,k=q\,,g\,,\bar q$ and $\alpha=0,\pm$.
The eigenvalues of the operator $J_{qg}$ give
the possible values of the spin in the quark-gluon channel and have
the form $j_{qg}=j_q+j_g+n=5/2+n$ with $n$ being a nonnegative integer.
The operator $J_{g\bar q}$ has a similar interpretation in the
gluon-antiquark channel. The $SL(2)$ symmetry of the evolution
kernels \re{ker-T} and \re{ker-S} becomes manifest
\be
[{\cal H}_{\cal F},L_\alpha]=[{\cal H}_{\cal F},L^2]=[L_\alpha,L^2]=0\,,
\label{sym}
\ee
where $L_\alpha$ ($\alpha=\pm,\,0$) are the total three-particle
$SL(2)$ generators
\be
L_\alpha = L_{\alpha,\bar q}+L_{\alpha,q} + L_{\alpha,g} \,,\qquad L^2 =
\frac12\left(L_+L_-+L_-L_+\right)+L_0^2\,. \label{L2}
\ee
Note that the conformal invariance holds separately for all terms
entering the $1/N_c-$expan\-sion of the evolution kernels \re{ker-S} and
\re{ker-T}.

Going from the matrix element of nonlocal operator to the quark-gluon
distribution function, Eq.~\re{dis-fun}, we find that
$D(x_1,x_2,x_3)$ obeys the RG equation similar to \re{RG} with the
Hamiltonian ${\cal H}_{\cal F}$ transformed from the
{\it coordinate\/} $z-$representation into the
{\it momentum\/} $x-$representation.
The analysis of the evolution equation for $D(x_1,x_2,x_3)$
is based on the solutions to the Schr\"odinger equation \cite{BDKM}
\be
{\cal H}_{\cal F}\cdot \Psi_{N,q}(x_1,x_2,x_3)
= {\cal E}_{N,q} \Psi_{N,q}(x_1,x_2,x_3)\,,
\label{H->Sch}
\ee
where the eigenstates $\Psi_{N,q}(x_i)$ are homogeneous polynomials in $x_i$ of
degree $N$ and index $q$ enumerates different energy levels
${\cal E}_{N,q}$. Here, the Hamiltonian is given by the evolution
kernels \re{ker-T} and \re{ker-S}
with the $SL(2;\mathbb R)$ generators defined in the
$x-$representation as
\ba
L_{k,0} \Psi(x_i) &=& (x_k\partial_k+j_k) \Psi(x_k)
\nonumber
\\
L_{k,+} \Psi(x_i) &=& -x_k \Psi(x_k)
\\
L_{k,-} \Psi(x_i) &=& (x_k\partial_{x_k}^2+2j_k\partial_{x_k})\Psi(x_k)\,.
\nonumber
\ea
Finally, solving the Schr\"odinger equation \re{H->Sch}
one can construct the basis of {\it local\/} conformal operators
\cite{BDKM}
\be
{\cal O}_{N,q}(0) =
\Psi_{N,q}(i\partial_{z_1},i\partial_{z_2},i\partial_{z_3})
{\cal F}(z_1,z_2,z_3)\bigg|_{z_i=0}\,,
\label{mult}
\ee
which have a fixed operator dimension, or equivalently a fixed total
number $N$ of covariant derivatives, and which do not mix under
renormalization. It follows from \re{H->Sch} that their anomalous
dimensions are determined by the energy of the state $\Psi_{N,q}$
\be
\gamma_q(N)=\frac{\alpha_s N_c}{2\pi}{\cal E}_{N,q}
\label{anom}
\ee
while their matrix elements over nucleon states evaluated at a
low normalization scale $\mu_0$ define the set of nonperturbative
dimensionless parameters
\be
\phi_{N,q}(\mu_0) = \vev {p,s |{\cal O}_{N,q}(0) |p,s}_{\mu=\mu_0}\,.
\ee
Substituting \re{dis-fun} into \re{mult} one finds the
solution to the evolution equation for (generalized) moments of
the distribution functions as
\be
\int_{-1}^1 {\cal D} x\,\Psi_{N,q}(x_1,x_2,x_3) {D}(x_1,x_2,x_3)
 =\phi_{N,q}(\mu_0)
\left(\frac{\alpha_s(\mu)}{\alpha_s(\mu_0)}\right)
^{{\cal E}_{N,q}/b_0}\,,
\label{mom}
\ee
where $b_0=11N_c/3-2n_f/3$ and ${\cal D}x=dx_1dx_2dx_3\delta(x_1+x_2+x_3)$.
Inverting this relation one can obtain the expression for the distribution
function as an integral over complex moments $N$.
\footnote{In the case of the twist-3 quark-gluon wave function,
$x_1+x_2+x_3=1$ and $x_i\ge 0$, the relation \re{mom} can be inverted
using the completeness condition for the set the states $\Psi_{N,q}(x_i)$
with $N={\rm nonnegative~integer}$ as \cite{BDKM}
$$
{D}(x_1,x_2,x_3;\mu) = x_1 x_2^2 x_3 \sum_{N\ge 0, q}
\phi_{N,q}
\Psi_{N,q}(x_i) \left(\frac{\alpha_s(\mu)}{\alpha_s(\mu_0)}\right)
^{{\cal E}_{N,q}/b_0}\,.
$$
}

Thus, the scale dependence of the twist-3 quark gluon distributions
becomes effectively equivalent to the eigenproblem \re{H->Sch}.
In what follows we shall solve exactly the Schr\"odinger equation
\re{H->Sch} and study in detail the properties of its solutions.
Analyzing \re{H->Sch} one has to identify the total set of the conserved
charges. The conformal symmetry allows to identify the degree $N$
of the polynomials $\Psi_{N,q}(x_i)$ as a trivial integral of motion.
For fixed $N$ the eigenstates $\Psi_{N,q}(x_i)$ belong to the irreducible
representation of the $SL(2,\mathbb{R})$ group parameterized by a total
conformal spin $h=N+j_{\bar q}+j_q+j_g$ and satisfy the relations
\ba
L_0 \Psi_h(x_i) &=& h\,
\Psi_h(x_i) \,,\quad h=N+\frac72
\nonumber
\\
L_- \Psi_h(x_i)&=&0
\label{h}
\\
L^2 \Psi_h(x_i) &=& h(h-1)
\Psi_h(x_i)\,.
\nonumber
\ea
Thus defined $\Psi_h(x_i)$ can be identified as the highest weights of
the $SL(2,\mathbb{R})$ representations labeled by a nonnegative integer $N$.
Note that for given $\Psi_h(x_i)$
there exists an infinite series of the ``excited'' eigenstates,
$(L_+)^m \Psi_h=(-x_1-x_2-x_3)^m\Psi_h$, which vanish due to
$x_1+x_2+x_3=0$.

However the conformal symmetry does not fix the eigenfunctions
$\Psi_h(x_i)$ uniquely. This can be seen in a number of ways. According to
\re{L2} the total conformal spin $h$ is equal to the sum of the conformal spin
of the quark-gluon system and the conformal spin of antiquark. For given $h$
the value of the spin in the $qg-$channel can be arbitrary,
$j_{qg}=j_q+j_g+m=5/2+m$ with $0 \le m \le N$, and this leads to an additional
degeneracy in solutions to \re{h}. The degeneracy can be removed if there
exists one more $SL(2)$ invariant conserved charge $Q$.  In this case, the
Schr\"odinger equation \re{H->Sch} becomes completely integrable -- the number
of degrees of freedom matches the number of the conserved charges, $L_0$, $L^2$
and $Q$, and, as a consequence, the wave functions, $\Psi_{N,q}$, and the
energy, ${\cal E}_{N,q}$, are uniquely fixed by the conformal spin $N$ and
the eigenvalues, $q$, of the charge $Q$.

It turns out that such operator exists for the evolution kernels
corresponding to the twist-3 quark-gluon operators in the multi-color limit,
Eqs.~\re{ker-T} and \re{ker-S}
\be
 [{\cal H}_{S^\pm}, Q_{S^\pm}]=0\,,\qquad [{\cal H}_T, Q_T]=0\,.
\ee
The $SL(2)$ invariant charges $Q_{S^+}$ and $Q_T$ are given by \cite{BDM}
\ba
Q_{S^+}&=& \{L_{12}^2,L_{23}^2\} - \frac12 L_{12}^2 - \frac92
L_{23}^2
\label{int-S}
\\
Q_T &=& \{L_{12}^2,L_{23}^2\} - \frac92 L_{12}^2 - \frac92
L_{23}^2\,,
\label{int-T}
\ea
where $\{\,,\}$ stands for an anticommutator, $L_{12}^2\equiv L_{qg}^2$ and
$L_{23}^2\equiv L_{g\bar q}^2$ are two-particle Casimir operators in the
quark-gluon and gluon-antiquark channels.
The charge $Q_{S^-}$ is obtained from $Q_{S^+}$ by permutation of
the particles $(123) \to (321)$. The kernel ${\cal H}_T$ is invariant
under this transformation and, as a consequence, its eigenstates
have a definite parity under $z_1\leftrightarrow z_3$.

\section{Integrability of the effective QCD Hamiltonian}

Let us demonstrate that the Schr\"odinger equation \re{H->Sch} defined
by the evolution kernels Eqs.~\re{ker-T} and \re{ker-S} is
equivalent in the multi-color limit to the one-dimensional Heisenberg
open spin chain model. To this end, we shall apply the Quantum Inverse Scattering
Method \cite{book,Skl} to construct an integrable (inhomogeneous) open spin chain
model and show that for certain values of parameters the Hamiltonian
of this model coincides with the QCD evolution kernels \re{ker-T}
and \re{ker-S}.

\subsection{Open spin chains}

To start with we consider $M$ particles with the
coordinates $x_n$ $(n=1,...,M)$ on a line and assign to each particle
three $SL(2)$ generators $(L_{+,n},L_{-,n},L_{0,n})$. In what
follows we shall refer to these generators as to spin$-j_n$ operators.
They are realized as differential operators \re{L's} acting on the
quantum space of the $n-$th particle that we denote as $V_{j_n}$.
The interaction between $M$ particles occurs through a pair-wise
interaction between the nearest-neighbor spins. The corresponding pair-wise
Hamiltonians depend on the two-particle Casimir operators $L_{jk}^2$
\be
L_{jk}^2=\sum_{\alpha=0,1,2}
(L_{\alpha,j}+L_{\alpha,k})^2\,,\qquad j\,,k=1\,,...\,,M
\label{L-2}
\ee
with $L_{\pm,j}\equiv L_{1,j}\pm iL_{2,j}$.

The definition of the model is based on
the existence of the solution to the Yang-Baxter equation
\be
R_{j_1j_2}(u) R_{j_1j_3}(v) R_{j_2j_3}(u-v)=
R_{j_2j_3}(u-v) R_{j_1j_3}(v) R_{j_1j_2}(u)\,.
\label{YB}
\ee
Here, the operator $R_{j_1j_2}(u)$ acts on the tensor product of
the quantum spaces $V_{j_1}\otimes V_{j_2}$ and depends on an
arbitrary complex spectral parameter $u$.
The solution to \re{YB} is given by \cite{KRS,XXX}
\be
R_{j_1j_2}(u)=(-)^{J_{12}+j_1+j_2}\frac{\Gamma(J_{12}-iu)}{\Gamma(J_{12}+iu)}
\frac{\Gamma(j_1+j_2+iu)}{\Gamma(j_1+j_2-iu)}\,,
\label{R}
\ee
where the operator $J_{12}$ is given by the sum of two $SL(2)$
spins and it is defined as a formal solution to
\be
J_{12}(J_{12}-1) = L_{12}^2\,.
\label{J12}
\ee
For arbitrary $SL(2)$ spins $j_1$ and $j_2$ the eigenvalues of the operator
$J_{12}$ have the form $j_{12}=j_1+j_2+n$ with $n=0\,,1\,,...$,
which correspond to decomposition of the tensor product $V_{j_1}\otimes V_{j_2}$ over
the irreducible $SL(2)$ components of spin $j_{12}$.

Having the explicit expression for the $R-$operator, Eq.\re{R}, we construct an integrable
open spin model following Sklyanin \cite{Skl}. Namely, we define the monodromy operator
acting on the space $V_j\otimes V_{j_1} \otimes ... \otimes V_{j_M}$
\be
T_{j}(u)=R_{j_1j}(u-i\omega_1) R_{j_2j}(u-i\omega_2) ...
R_{j_Mj}(u-i\omega_M)
\label{T}
\ee
with $j$ being the spin of the auxiliary space $V_j$ and
$\omega_k$ being the shifts (spin chain impurities) associated with
the $k-$th particle. Let us impose the additional condition that
the spins and the impurities of all ``intermediate'' sites
are the same
\be
j_2=j_3=...=j_{M-1}\,,\qquad
\omega_2=\omega_3=...=\omega_{M-1}\,,
\label{para}
\ee
while $j_2\neq j_1,j_M$ and $\omega_2\neq \omega_1,\omega_M$.

As we will show in Sect.~3.2 it is this condition (together
with $\omega_2=0$ and $M=3$) that one finds matching the QCD evolution
kernels, \re{ker-T} and \re{ker-S}, into integrable spin chain Hamiltonian.
Then, the transfer matrix of the open spin chain with $M$ sites is defined as
\footnote{General definition of integrable open spin chain models involves the boundary
matrices $K_\pm(u)$ satisfying the reflection Yang-Baxter equation
\cite{Ch,Skl,KS}.
We have chosen the simplest solution $K_\pm=\mathbb{I}$ to ensure the $SL(2)$
invariance of the Hamiltonian, Eq.\re{sym}.}
\be
t_j(u)=\tr_j \left[ T_j(u) T^{-1}_j(-u)\right]\,.
\label{t}
\ee
Here, the trace is taken over the auxiliary $SL(2)$ representation space of the spin
$j$. Choosing different values of $j$ one obtains (an infinite) set of the transfer
matrices $t_j(u)$ which by virtue of the Yang-Baxter equation \re{YB} commute
with each other for different values of the spectral parameters
\be
 [t_{j_1}(u), t_{j_2}(v)]=0
\label{mut}
\ee
for any $j_1$ and $j_2$. As a consequence, expanding $t_j(u)$ in powers of the
spectral parameter $u$ and choosing different values of the spin $j$ one obtains the
family of mutually commuting operators. Their explicit form
depends on the set of parameters, $j_k$ and $\omega_k$, which define
the spectrum of spins and impurities
of the model, respectively. This family contains
the Hamiltonian ${\cal H}_M$ of the model as well as a complete set of the conserved
charges.
In addition, the transfer matrix possesses an additional
$SL(2)$ symmetry%
\footnote{Indeed, it follows from the definition \re{T} that the monodromy
operator commutes with the sum of the total spin $\vec L$ and
the auxiliary spin $\vec L_j$, $[T_j(u), \vec L_j + \vec L]=0$.
To get \re{t-sym} one has to take the trace over $V_j$ and use its cyclic
symmetry property.}
\be
[t_j(u), L_\alpha]=0\,,\qquad L_\alpha=L_{\alpha,1}+ ... +
L_{\alpha,M}\,,\quad \alpha=0\,,\pm\,,
\label{t-sym}
\ee
which ensures the $SL(2)$ invariance of the Hamiltonian, \re{sym}.

\subsection{Integrals of motion}

Let us show that the QCD integrals of motion \re{int-T} and \re{int-S}
appear in the expansion of the spin$-1/2$ transfer matrix, $t_{-1/2}(u)$,
given by \re{t} with the auxiliary $SL(2)$ spin%
\footnote{Throughout this paper we shall use the convention
$j_{_{SU(2)}}=-j_{_{SL(2)}}$.}
$j=-1/2$ and the number
of sites $M=3$.%

It is well known \cite{F,book}, that the spin$-1/2$ $R-$operator \re{R}
coincides with the Lax operator of the Heisenberg spin chain,
$R_{-\frac12j_k}(u)\equiv {\mathbb L}_{j_k}(u)$
\be
{\mathbb L}_j(u)= u + \frac{i}2 - i\sum_{\alpha=0,\pm} L_\alpha\,
\sigma^\alpha =\matrix{u+\frac{i}2-iL^3 &  -i L^- \\ -i L^+ &
u+\frac{i}2+i L^3}
\label{Lax}
\ee
with $L_\alpha$ being spin$-j$
generators and $L_\pm=L_1\pm iL_2$. Substituting \re{Lax} into
\re{T} one evaluates the spin$-1/2$ monodromy operator as a $2\times
2$ matrix
\be
T_{-1/2}(u)= {\mathbb L}_{j_1}(u-i\omega_1) ... {\mathbb L}_{j_M}(u-i\omega_M)
=\matrix{ a(u) & b(u) \\ c(u) & d(u) }
\,.
\label{abcd}
\ee
Here, $a(u)$, $b(u)$, $c(u)$ and $d(u)$ are expressed in
terms of the spin operators and satisfy the Yang-Baxter relations of the
form
\be
 [b(u),c(v)]=i\frac{d(v)a(u)-a(v)d(u)}{u-v}
\label{bc}
\ee
together with similar relations for other components of
$T_{-1/2}(u)$. The spin$-1/2$ transfer matrix $t_{-1/2}(u)$ is given by \re{t}
with $T_{-1/2}(u)$ replaced by its expression \re{abcd}. It proves
convenient to change the normalization of the transfer matrix
by introducing the following operator \cite{Skl}
\be
\widehat t_{-1/2}(u) = t_{-1/2}\left(-u-\frac{i}2\right)
\rho(u)\,,
\label{rho-half}
\ee
where the c-valued factor $\rho(u)$ compensates the poles of
$t_{-1/2}(-u-i/2)$ and is given by
\be
\rho(u)=\prod_{k=1}^M (u+ij_k^-)(u-ij_k^++i)
\ee
with the parameters $j_k^\pm$ defined as
\be
j_k^\pm=j_k\pm \omega_k \,.
\label{j-pm}
\ee
Thus defined normalized transfer matrix can be expressed as
\cite{Skl}
\be
\widehat t_{-1/2}(u)=\tr\, \widehat T_{-1/2}(u)\,,\qquad
\widehat T_{-1/2}(u)
= T_{-1/2}\lr{u-\frac{i}2}\, \sigma_2\,
T_{-1/2}^t\lr{-u-\frac{i}2}\, \sigma_2
\label{P3}
\ee
and it has the following properties. Firstly, in contrast with
$t_{-1/2}(u)$, the normalized transfer matrix does not have poles in
$u$ and, secondly, $\widehat t_{-1/2}(u)$ has the form of a polynomial of degree
$M$ in $u^2$ with the corresponding operator coefficients providing the
set of the conserved charges of the model.

It becomes straightforward to calculate the normalized transfer
matrix $\widehat t_{-1/2}(u)$ for the chain with $M=3$ sites. Substituting \re{Lax}
into \re{abcd} and \re{P3} one finds after some algebra%
\footnote{Similar calculation of the transfer matrix has been performed in the
second publication in \cite{B}.
We disagree with Eq.~(47) there in the coefficient in front of $\tau(u^2)$.}
\be
\widehat t_{-1/2}(u)= (4u^2+1) \tau(u^2) - 2\prod_{k=1,2,3} \left(u^2-L_k^2
+\omega_k^2\right)
\label{t-hat}
\ee
with one-particle Casimir operators
$L_k^2=j_k(j_k-1)$ being c-numbers and the operator $\tau(u^2)$ defined as
\ba
\tau(u^2)&=& u^2[L^2 - \sum_{k=1,2,3} L_k^2]
         + L^2 \left[ L_2^2+\omega_2^2 \right]
         - \frac12 Q
\nonumber
\\
         && + L_1^2 L_3^2
         -L_1^2\omega_3^2
         -L_2^2(\omega_1^2-\omega_2^2+\omega_3^2)
         -L_3^2\omega_1^2\,.
\ea
The charge $Q$ is given by
\be
Q= - 2 {\omega_2}\ [L_{12}^2,L_{23}^2]_-
  +  \{L_{12}^2,L_{23}^2\}_+
  + 2 (\omega_2^2-\omega_3^2)\ L_{12}^2
  + 2 (\omega_2^2-\omega_1^2)\ L_{23}^2\,.
\label{Q}
\ee
with $L_{jk}^2$ and $L^2$ being two- and three-particle Casimir
operators defined in \re{L-2} and \re{L2}, respectively.

The normalized transfer matrix \re{t-hat} involves two mutually commuting
operators, $Q$ and $L^2$. Together with $L_0$ they form the complete set of the
conserved charges for integrable open chain of $M=3$ particles with arbitrary
values of the spins $j_k$ and the impurity parameters $\omega_k$. To fix
their values we compare \re{Q} with the expressions for the integrals of motion
of the QCD evolution equations, Eqs.~\re{int-T} and \re{int-S}.
We find that the spins of particles are equal to the conformal spins of the
corresponding fields
\be
j_1=j_3=1\,,\qquad j_2=\frac32\,,
\label{j's}
\ee
and the impurity parameters are given by
\be
Q_{S^+}:\qquad
\omega_1^2=\frac94\,,\quad\omega_2=0\,,\quad \omega_3^2=\frac14
\label{wS}
\ee
and
\be
Q_{T}\ :\qquad
\omega_1^2=\frac94\,,\quad\omega_2=0\,,\quad \omega_3^2=\frac94\,.
\label{wT}
\ee
These matching conditions define the parameters $\omega_1$ and $\omega_3$ up
to a sign. As we will show in Sect.~3.3, the Hamiltonian of the model is
invariant under $\omega_{1,3}\to -\omega_{1,3}$.

Eqs.~\re{j's}, \re{wS} and \re{wT} establish the equivalence relations
between the QCD evolution kernels and the open spin chain models.
Instead of considering two cases \re{wS} and \re{wT} separately, we
shall treat $\omega_1$ and $\omega_3$ as free parameters and
introduce the charge
\be
Q(\omega_1,\omega_3)=  \{L_{12}^2,L_{23}^2\}_+
     - 2 \omega_3^2 \ L_{12}^2 - 2 \omega_1^2 \ L_{23}^2\,.
\label{Q(a)}
\ee
In what follows we shall find the flow of the eigenvalues of this
charge, $q(\omega_1,\omega_3)$, in the parameters $\omega_{1,3}$ and identify the
integrals of motion, $q_T$ and $q_{S^+}$, as
corresponding to the special values of the flow parameters
\be
q_{S^+}=q\left(\frac32,\frac12\right)\,,\qquad
q_{T}=q\left(\frac32,\frac32\right)\,.
\label{range}
\ee

\subsection{Integrable Hamiltonian}

Let us complete an identification of the QCD evolution equations as
integrable systems by showing that the evolution kernels \re{ker-T} and
\re{ker-S} coincide with the Hamiltonian of open spin chain model.
The latter enters into the expansion of the transfer matrix $t_j(u)$ given
by \re{t} with the auxiliary spin $j$ equal to the spin of intermediate
``gluonic'' site $j=j_2=3/2$. The corresponding monodromy operator takes the form
\be
T_{j}(u) = R_{j_1j}(u-i\omega_1) R_{j_2j}(u) R_{j_3j}(u-i\omega_3)
\bigg|_{j=j_2}\,,
\label{T-QCD}
\ee
where $\omega_2=0$, $j_1=j_3=1$ and $j_2=3/2$.

We would like to stress that the spin chain described by \re{T-QCD} is
inhomogeneous - the spins and impurities of the end-points are different
from the ones of the internal sites of the chain. To our best knowledge such
QCD induced spin magnets have not been studied before and their analysis
represents a certain interest from point of view of integrable models. It
is for this reason that we generalize the definition
\re{T-QCD} by enlarging the size of the open spin chain to an arbitrary
number of sites, $M\ge 3$, and fixing the parameters (spins and impurities)
according to \re{para}
\be
T_{j}(u) = R_{j_1j}(u-i\omega_1) R_{j_2j}(u)...R_{j_{M-1}j}(u)
R_{j_Mj}(u-i\omega_M) \bigg|_{j=j_2=...=j_{M-1}}\,,
\label{T-N}
\ee
where we put $\omega_2=...=\omega_{M-1}=0$ in order to match \re{T-QCD}
for $M=3$.

Finally, one finds the transfer matrix $t_{j}(u)$ by substituting the monodromy operator
\re{T-N} into \re{t} and defines the Hamiltonian of inhomogeneous spin chain with
$M$ sites as
\be
{\cal H}_M = \frac{i}2 \partial_u \ln t_{j}(u=0)
= \frac{i}2 \partial_u \ln
\tr\left[T_{j}(u) T_{j}^{-1}(-u)\right]\bigg|_{u=0}
\,.
\label{H}
\ee
We would like to notice that this relation differs from the
standard definition of Hamiltonian based on the so-called fundamental
monodromy operator \cite{Skl,F}. In contrast with the latter, the
spins of quantum spaces $V_{j_1}$ and $V_{j_M}$
corresponding to the end-points of the chain \re{T-N} are different from
that of the auxiliary space $V_j$.
\footnote{At this point we disagree with the statement made in
\cite{B} that the Hamiltonian \re{H-open} arises from the
expansion of the fundamental transfer matrix.}

One finds the explicit form of the Hamiltonian \re{H}
in the standard way \cite{Skl}
by replacing the monodromy operator $T_j(u)$ by its expression
\re{T-N} and taking into account the following property of the
$R-$operator \re{R}
\be
R_{j_kj}(0)= {\mathbb P}_{j_kj}\,,\qquad k=2,...,M-1 \,,
\label{perm}
\ee
where ${\mathbb P}_{j_2j}$ is the permutation operator acting on the tensor
product of the quantum spaces as
${\mathbb P}_{j_2j}[V_{j_2}\otimes V_j]=V_j\otimes V_{j_2}$ \cite{KRS,XXX}.
Note that \re{perm} holds only for intermediate sites with $j_k=j$
while the operators $R_{j_1j}$ and $R_{j_Mj}$ corresponding to the end-points of
the chain do not possess such property. As a consequence, the calculation of the
Hamiltonian \re{H} deviates the standard derivation \cite{Skl} and requires a special
consideration.

Substituting \re{T-QCD} into \re{t} and \re{H} and making use of \re{perm} we obtain
after some algebra
\be
{\cal H}_M= H_{j_1j_2}(\omega_1) +
\sum_{k=2}^{M-2}H_{j_kj_{k+1}}(0) +
H_{j_{M-1}j_M}(\omega_M) + \Delta {\cal H}\,,
\label{H-open}
\ee
where the notation was introduced for the pair-wise Hamiltonian
\be
H_{j_kj_{k+1}}(\omega)= -\partial_\omega R_{j_kj_{k+1}}(-i\omega)
R_{j_kj_{k+1}}^{-1}(-i\omega)
=-\partial_\omega R_{j_kj_{k+1}}(-i\omega)
R_{j_kj_{k+1}}(i\omega)
\label{pair}
\ee
and the operator $\Delta {\cal H}$ is defined as
\be
\Delta {\cal H} = \tr_j \left\{
H_{j_1j}(\omega_1) + R_{j_1j}(-i\omega_1) H_{j_2j}(0)R_{j_1j}(i\omega_1)
-H_{j_1j_2}(\omega_1)
\right\} / \tr_j \II\,.
\label{dH}
\ee
In comparison with the definition of the homogeneous spin chain Hamiltonian \cite{Skl},
the expression \re{H-open} contains the additional operator $\Delta{\cal
H}$. One can show however that $\Delta{\cal H}$ provides a c-number
correction to ${\cal H}_M$. To this end one differentiates
the both sides of the Yang-Baxter equation
$$
R_{jj_1}(u-i\omega_1)R_{j_2j}(u)R_{j_2j_1}(-i\omega_1)
=R_{j_2j_1}(-i\omega_1)R_{j_2j}(u)R_{jj_1}(u-i\omega_1)
$$
with respect to the spectral parameter $u$ and puts $u=0$. Then, multiplying
the both sides of the relation from the right by ${\mathbb P}_{j_2j}$ and
using \re{perm} one arrives at
\be
H_{jj_1}(\omega_1) + R_{jj_1}(-i\omega_1) H_{j_2j}(0) R_{jj_1}(i\omega_1)
=
H_{j_2j_1}(\omega_1) + R_{j_2j_1}(\omega_1) H_{j_2j}(0)R_{j_2j_1}(-i\omega_1)\,.
\ee
Using this identity one evaluates \re{dH} as%
\footnote{Here, the last relation in the r.h.s.\ follows from the $SL(2)$
invariance of the 2-particle Hamiltonian,
$\tr_j [L_{j_2}^\alpha+L_j^\alpha,H_{j_2j}(0)]=[L_{j_2}^\alpha,\tr_j
H_{j_2j}(0)]=0$.
}
\be
\Delta {\cal H}
=R_{j_1j_2}(i\omega_1)\frac{\tr_j H_{j_2j}(0)}{\tr_j \II}
 R_{j_1j_2}(-i\omega_1)
=\frac{\tr_j H_{j_2j}(0)}{\tr_j \II} = {\rm const}\times \II_{j_2}\,.
\ee
In what follows we shall neglect the additive correction due to $\Delta {\cal H}$.

Thus, the integrable Hamiltonian of the inhomogeneous open spin chain with $M$
sites is given by \re{H-open} with the pair-wise Hamiltonian
defined by \re{pair} and \re{R} as
\be
H_{j_1j_2}(\omega)=\psi(j_{12}-\omega) + \psi(j_{12}+\omega)
                  -\psi(j_1+j_2-\omega)- \psi(j_1+j_2+\omega)
\label{H12}
\ee
with $\psi(x)=d\ln\Gamma(x)/dx$ and the operator $j_{12}$ introduced in
\re{J12}. This expression is valid for arbitrary spins $j_1$, $j_2$ and,
as it was anticipated, it is an even function of the impurity parameter
$\omega$.

It is now straightforward to see that for $M=3$ and the parameters
$j_k$ and $\omega_k$ given by \re{j's}, \re{wS} and \re{wT}, the
Hamiltonian \re{H-open} coincides (up to an overall normalization) with the
QCD evolution kernels \re{ker-T} and \re{ker-S}. Following \re{Q(a)} we
introduce the interpolating Hamiltonian depending on the flow
parameters $\omega_{1,3}$
\ba
{\cal H}(\omega_1,\omega_3) &=&
                    \psi(j_{12}-\omega_1)
                   +\psi(j_{12}+\omega_1)
                   -\psi(\mbox{$\frac52$}-\omega_1)
                   -\psi(\mbox{$\frac52$}+\omega_1)
\nonumber
\\
                   &+&\psi(j_{23}-\omega_3)
                   +\psi(j_{23}+\omega_3)
                   -\psi(\mbox{$\frac52$}-\omega_3)
                   -\psi(\mbox{$\frac52$}+\omega_3)\,.
\label{H(omega)}
\ea
Then, the relation between ${\cal H}(\omega_1,\omega_3)$ and the QCD evolution
kernels, Eqs.~\re{ker-T} and \re{ker-S}, looks like
\be
{\cal H}_{S^+}= {\cal H}
                  \left(\frac32,\frac12\right)+ \frac{17}{6}\,,\qquad
{\cal H}_{T}= {\cal H}
                  \left(\frac32,\frac32\right)+ \frac{13}{6}\,.
\label{const}
\ee
Thus, the energies ${\cal E}_{S^\pm}$ and ${\cal E}_T$ entering
into the solutions to the QCD evolution equations \re{mom}
for the twist-3 quark-gluon distributions, $D_{S^\pm}(x_i)$
and $D_T(x_i)$, respectively, can be calculated as
\be
{\cal E}_{S^+}= {\cal E}
                  \left(\frac32,\frac12\right)+ \frac{17}{6}
                  +{\cal O}(1/N_c^2)\,,\qquad
{\cal E}_{T}= {\cal E}
                  \left(\frac32,\frac32\right)+ \frac{13}{6}
                  +{\cal O}(1/N_c^2)\,,
\ee
where ${\cal E}(\omega_1,\omega_3)$ are eigenvalues of the Hamiltonian
\re{H(omega)} and $1/N_c^2-$corrections correspond to nonplanar
part of the evolution kernels \re{ker-T} and \re{ker-S}.

Having identified the QCD evolution kernels \re{ker-T} and \re{ker-S} as
Hamiltonians of the open spin chain model we now turn to the eigenvalue problem
for \re{H-open}. The complete integrability of the Hamiltonian \re{H-open}
implies that ${\cal H}_M$ is a (complicated) function of
the conserved charges. Using this dependence to which we shall
refer as to the dispersion curve we can find the spectrum of the
Hamiltonian by replacing the charges by their corresponding
eigenvalues. However, instead of considering separately
the eigenproblem for
${\cal H}_M$ we shall solve a more general problem of diagonalization
the transfer matrices $t_j(u)$ of different spins $j$.

We recall that
the conserved charges and the Hamiltonian are generated by two different
transfer matrices, $t_{-1/2}(u)$ and $t_{j_2}(u)$, respectively.
As we will show in Sect.~4, the transfer matrix $t_{-1/2}(u)$
can be diagonalized using the Algebraic Bethe Ansatz (ABA) \cite{F}. This
immediately gives the eigenvalues of the integrals of motion.
Moreover, as we will argue in Sect.~5, the transfer matrices of
higher spins, $t_j(u)$, are related to $t_{-1/2}(u)$ through
the nonlinear recurrence relations -- the so-called fusion hierarchy
\cite{KRS,KR,MN,Zhou,BPO}.
Together with the ABA expressions for eigenvalues of $t_{-1/2}(u)$
these relations allow to reconstruct the spectrum of $t_{j}(u)$
and finally solve the eigenproblem for ${\cal H}_M$.

\section{Diagonalization of the conserved charges}

Let us find the spectrum of the transfer matrix $\widehat t_{-1/2}(u)$.
Throughout this section we shall keep the number of sites $M$
arbitrary. According to the definition \re{t-hat}, $\widehat t_{-1/2}(u)$ is an
operator acting on the quantum space of $M$ particles,
$V_{j_1}\otimes ...\otimes V_{j_M}$. The quantum space of the $k-$th
particle, $V_{j_k}$ has the highest weight
\begin{equation}
\ket{0_k}=1\,,\qquad
L_{-,j_k} \ket{0_k} = 0\,,\qquad
L_{0,j_k} \ket{0_k} = j_k \ket{0_k}\,.
\end{equation}
Combining together the highest weights
of $M$ particles we construct the pseudovacuum state
\be
\ket{\Omega_+} = \prod_{k=1}^M\ket{0_{j_k}}=1
\,,\quad L_-\ket{\Omega_+} =0\,,\quad
L_0\ket{\Omega_+} = (j_1+..+j_M)\ket{\Omega_+}
\label{pseudo}
\ee
with $L_\alpha$ being the total spin operator defined in \re{L2}.
As we will see in a moment, $\ket{\Omega_+}$ is the eigenstate
of the Hamiltonian \re{H-open} with the spin $N=0$.

\subsection{Algebraic Bethe Ansatz}

The existence of the pseudovacuum state \re{pseudo} allows
to apply the Algebraic Bethe Ansatz (ABA) for diagonalization of
the transfer matrix $t_{-1/2}(u)$. Following the standard procedure
\cite{F,Skl}, we first examine the action of the spin$-1/2$ monodromy operator
$T_{-1/2}(u)$ on the pseudovacuum state \re{pseudo}.

Due to the definition \re{T}, $T_{-1/2}(u)$ is given by a product of the Lax
operators \re{Lax}. Acting on the pseudovacuum state \re{pseudo}
the Lax operators take the triangle form
\be
{\mathbb L}_{j_k}(u-i\omega_k) \ket{\Omega_+} =
\matrix{u-i\omega_k+\frac{i}2 -i j_k & 0 \\ {*} & u-i\omega_k+\frac{i}2 +i j_k }
\ket{\Omega_+}\,.
\ee
This allows to calculate their product in Eq.\re{T} as
\be
a(u)\ket{\Omega_+}=\delta_+\lr{u+\frac{i}2} \ket{\Omega_+}\,,
\quad
b(u)\ket{\Omega_+}=0\,,
\quad
d(u)\ket{\Omega_+}=\delta_-\lr{u+\frac{i}2} \ket{\Omega_+}\,,
\label{acd}
\ee
where the operators $a(u)...d(u)$ were defined in \re{abcd} as
different components of $T_{-1/2}(u)$ and the notation was introduced
for the functions
\be
\delta_\pm(u)=\prod_{k=1}^M (u-i(\omega_k\pm j_k))\,.
\ee
Substituting \re{abcd} into \re{P3} one gets
\be
\widehat T_{-1/2}(u) = \tr \matrix{ A(u) & B(u) \\ C(u) & D(u)}
\,,\qquad
\widehat t_{-1/2}(u)
= A(u)+D(u)\,,
\label{A+D}
\ee
where the operators $A(u)$ and $D(u)$ are defined as
\ba
A(u)&=&a\lr{u-\frac{i}2}d\lr{-u-\frac{i}2}
-b\lr{u-\frac{i}2}c\lr{-u-\frac{i}2}
\nonumber
\\
D(u)&=&d\lr{u-\frac{i}2}a\lr{-u-\frac{i}2}
-c\lr{u-\frac{i}2}b\lr{-u-\frac{i}2}
\label{AD}
\ea
and the explicit form of the operators $C(u)$ and $B(u)$ is not
relevant for our purposes. Using \re{acd} together with \re{bc} we
find that the pseudovacuum diagonalizes both operators
\ba
A(u)\ket{\Omega_+}&=&\delta_+(u)\delta_-(-u) \ket{\Omega_+}
\\
D(u)\ket{\Omega_+}&=&\left\{\lr{1-\frac{i}{2u}}\delta_+(-u)\delta_-(u)
+\frac{i}{2u}\delta_+(u)\delta_-(-u)
\right\}\ket{\Omega_+}\,.
\ea
and, as a consequence, we obtain from \re{A+D}
\be
\widehat t_{-1/2}(u) \ket{\Omega_+}
=\left\{\frac{2u+i}{2u}\delta_+(u)\delta_-(-u)
       +\frac{2u-i}{2u}\delta_+(-u)\delta_-(u)
 \right\}\ket{\Omega_+}\,.
\label{t-0}
\ee
This relation implies that the pseudovacuum state is an eigenstate of the
Hamiltonian \re{H-open}. Comparing \re{pseudo} with \re{h} we identify
$\ket{\Omega_+}$ as the eigenstate with the minimal conformal
spin, $N=0$,
\be
\Psi_{N=0}(x_1,...,x_M) =\ket{\Omega_+}=1\,.
\label{BS0}
\ee
To find the corresponding eigenvalues of the conserved charges
one has to expand the r.h.s.\ of \re{t-0} in powers of $u^2$.
In particular, for $M=3$ and the parameters $j_k$ and $\omega_k$
given by \re{j's} and \re{Q(a)}
we match \re{t-0} into \re{P3} to obtain
the corresponding eigenvalue of the integral of motion as
\be
Q(\omega_1,\omega_3) \ket{\Omega_+} = \frac{15}2\left( \frac{15}4
-\omega_1^2-\omega_3^2 \right) \ket{\Omega_+}\,.
\ee
One can also get this relation using the definition \re{Q(a)} and taking
into account that $L_{ik}^2|\Omega_+\rangle = (j_i+j_k)(j_i+j_k-1)
|\Omega_+\rangle$. The same identity allows to calculate the
corresponding energy \re{H(omega)} as
\be
{\cal H}(\omega_1,\omega_3) \Psi_{N=0}(z_1,...,z_M)=0\,,
\label{zero-en}
\ee
or equivalently ${\cal E}(\omega_1,\omega_3)=0$ at $N=0$.

For higher values of the conformal spin $N\ge 1$ the eigenstates
of the auxiliary transfer matrix, or equivalently the eigenstates
of the Hamiltonian ${\cal H}_M$, are given by the Bethe States
\cite{F,Skl}
\be
\Psi_N(x_1,...,x_M)=
\ket{\lambda_1,...,\lambda_N} = C(\lambda_1)... C(\lambda_N)
\ket{\Omega_+}
\label{BS}
\ee
and are parameterized by the set of (complex) parameters
$\{\lambda_k\}$. Here, $C(u)$ is off-diagonal element of the
monodromy operator defined in \re{A+D}. It becomes straightforward
to verify that these states satisfy the highest weight condition
\re{h} and diagonalize the transfer matrix provided that the
complex parameters $\{\lambda_j\}$ satisfy certain conditions to
be specified below in Eq.~\re{BE}. The corresponding eigenvalues of the
normalized transfer matrix are given by
\ba
\widehat t_{-1/2}(u)&=&\frac{2u+i}{2u}\delta_+(u)\delta_-(-u)
\prod_{j=1}^N \frac{(\lambda_j+u-i)(\lambda_j-u+i)}
{(\lambda_j+u)(\lambda_j-u)}
\nonumber
\\
&+&\frac{2u-i}{2u}\delta_+(-u)\delta_-(u)
\prod_{j=1}^N \frac{(\lambda_j+u+i)(\lambda_j-u-i)}
{(\lambda_j+u)(\lambda_j-u)}\,.
\label{t-M}
\ea
They differ from \re{t-0} by two ``dressing'' factors
depending on the Bethe roots $\{\lambda_j\}$ and having poles at $u=\pm \lambda_j$.
Since according to the definition \re{P3}, $\widehat t_{-1/2}(u)$ is an
even polynomial in $u$ of degree $2M$, the residue of the
r.h.s.\ of \re{t-M} at $u=\pm \lambda_j$ and $u=0$ should vanish.
One verifies that the pole at $u=0$ cancels in the sum of two terms
\re{t-M}. The condition for $\widehat t_{-1/2}(u)$ to have zero residue at
$u=\lambda_k$ leads to the set of the Bethe equations on
the complex parameters $\{\lambda_j\}$
\be
\prod_{j=1}^N
\frac{(\lambda_k+\lambda_j+i)(\lambda_k-\lambda_j+i)}
     {(\lambda_k+\lambda_j-i)(\lambda_k-\lambda_j-i)}
=\frac{\Delta_+(\lambda_k)}{\Delta_-(\lambda_k)}\,,
\label{BE}
\ee
where the notation was introduced for the functions
\ba
\Delta_+(u)&=&\frac{2u+ i}{2u} \delta_+(u)\delta_-(-u)
=(-1)^M \frac{2u+i}{2u} \prod_{k=1}^M
(u-ij_k^+) (u-ij_k^-)
\label{Delta}
\\
\Delta_-(u)&=&\Delta_+(-u)
\nonumber
\ea
with the parameters $j_k^\pm$ given by \re{j-pm}.

\subsection{Baxter ${\mathbf Q}-$function}

To calculate the spectrum of the transfer matrix \re{t-M}
one has to find all possible solutions $\lambda_1$, $...$, $\lambda_N$
to the Bethe Equation \re{BE} for different values of the conformal spin $N$.
Unfortunately, there exist no regular way of solving the Bethe
Equations \re{BE}. A more efficient (although equivalent) way of
calculating the transfer matrix is based on the Baxter
${\mathbf Q}-$function \cite{Bax}.

Let us define the Baxter ${\mathbf Q}-$function as an even polynomial in the
spectral parameter $u$ of degree $2N$ with the zeros given by the Bethe
roots
\be
{\mathbf Q}(u)=\prod_{j=1}^N (u+\lambda_j)(u-\lambda_j)\,.
\label{Q-roots}
\ee
Then, the obtained expression for the transfer matrix, \re{t-M},
is equivalent to the following 2nd order finite-difference equation
on ${\mathbf Q}(u)$
\be
\widehat t_{-1/2}(u)\ {\mathbf Q}(u)= \Delta_+(u)\
{\mathbf Q}(u-i) + \Delta_-(u)\ {\mathbf Q}(u+i)
\label{Bax}
\ee
with $\Delta_\pm(u)$ given by \re{Delta}. Solving this equation
one is interesting in finding all possible {\it polynomial\/}
solutions for ${\mathbf Q}(u)$. Each polynomial solution to
the Baxter equation is in one-to-one correspondence with
the Bethe states \re{BS} -- the degree of the polynomial
defines the conformal spin of the states and its zeros
provide the Bethe roots. In particular, the pseudovacuum
state \re{BS0} is related to a trivial solution to the
Baxter equation
\be
{\mathbf Q}_{N=0}(u) =1\,.
\label{Q-tri}
\ee
We shall study the properties of the
Baxter equation in more detail in Sects.~6 and 7.

Thus, applying the Algebraic Bethe Ansatz we were able to diagonalize the
transfer matrix $t_{-1/2}(u)$. Expanding \re{t-M} in powers of $u$ we can
reconstruct the spectrum of the integrals of motion. The next step
should be diagonalization of the transfer matrix $t_{j}(u)$ generating
the Hamiltonian \re{H-open}. We recall that the only difference between
$t_{-1/2}(u)$ and $t_{-j}(u)$ is in the spin of the auxiliary space
$V_j$ entering the definition \re{t}. It is well known that
higher (integer or half-integer) spin$-j$ representations can be obtained
from $2j$ copies of spin$-1/2$ representations through the so-called
fusion procedure. Applying the same procedure to the transfer matrix
of higher spin and performing fusion in the auxiliary space
\cite{KRS} one can express $t_{-j}(u)$ in terms of spin$-1/2$ transfer
matrices $t_{-1/2}(u)$ and then generalize it to arbitrary $j$.
Such relations -- the so-called $SU(2)$ fusion hierarchy
-- were first derived for the closed spin chains \cite{KR} and later
generalized to the open spin chains \cite{MN,Zhou,BPO}.

\section{Fusion hierarchy}

The fusion hierarchy provides the set of nonlinear recurrence relations
between the transfer matrices of different spins. These relations have the
same form for the closed and open spin chains \cite{Zhou}%
\footnote{Our definition of the $SL(2)$ transfer matrix is related
to that used in \cite{Zhou} as $\widehat t_{-j}=T^{(2j)}_0$.}
\be
\widehat t_{-j}(u)\, \widehat t_{-1/2}(u+2ij) = \widehat t_{-j-1/2}(u)
+ \widehat t_{-j+1/2}(u)\, f(u+(2j-1)i)\,,\qquad
\widehat t_0(u)=1\,,
\label{fusion}
\ee
but the explicit form of the c-valued function $f(u)$ is different
in two cases. The fusion relation \re{fusion} is a direct consequence
of the spin addition rule and the
shifts of the spectral parameter in the arguments of $\widehat t_{-1/2}$
and $f$ are due to the fact that transfer matrices are built from
noncommutative $R-$operators. Here, $\widehat t_{-j}(u)$ is the normalized
transfer matrix of the $SU(2)$ spin $j$
\be
\widehat t_{-j}(u) \equiv t_{-j}(-u-ij) \rho_j(u)\,,
\label{tj-hat}
\ee
which differs from \re{t} by a multiplicative c-valued factor
\be
\rho_j(u) = \prod_{k=0}^{2j-1} \rho(u+ik)
\ee
that compensates all poles of $t_{-j}(u)$. Similar to $\widehat t_{-1/2}(u)$,
Eqs.~\re{rho-half} -- \re{t-hat}, the normalized transfer matrix
$\widehat t_{-j}(u)$ is given by a polynomial in $u$ of degree $4Mj$ with
operator valued coefficients.

To specify the function $f(u)$ entering \re{fusion} let us
supplement the fusion relations \re{fusion} by the expression
for $\widehat t_{-1/2}(u)$ in terms of the Baxter ${\mathbf Q}-$function
\re{Bax}
\be
\widehat t_{-1/2}(u)= \Delta_+(u)\ \frac{{\mathbf Q}(u-i)}{{\mathbf Q}(u)}
               + \Delta_-(u)\ \frac{{\mathbf Q}(u+i)}{{\mathbf Q}(u)}\,.
\label{t-Q}
\ee
Then, applying \re{fusion} and \re{t-Q} one can find
the transfer matrices of higher spins through a single ${\mathbf Q}-$function.
In particular, for $j=-1/2$ one obtains from \re{fusion} the
spin$-1$ transfer matrix as
\be
\widehat t_{-1}(u)=\widehat t_{-1/2}(u)\,\widehat t_{-1/2}(u+i) - f(u)
\label{t-Q1}
\ee
and taking into account \re{t-Q} one gets
\ba
\widehat t_{-1}(u)&=&{\mathbf Q}(u-i){\mathbf Q}(u+2i) \left\{
\frac{\Delta_-(u)\Delta_-(u+i)}{{\mathbf Q}(u-i){\mathbf Q}(u)}
+\frac{\Delta_+(u)\Delta_-(u+i)}{{\mathbf Q}(u){\mathbf Q}(u+i)}
+\frac{\Delta_+(u)\Delta_+(u+i)}{{\mathbf Q}(u+i){\mathbf Q}(u+2i)}
\right\}
\nonumber
\\
&&+\delta\widehat t_{-1}(u)
\ea
with $\delta\widehat t_{-1}(u)=\Delta_-(u)\Delta_+(u+i)-f(u)$. Then,
the function $f(u)$ can be fixed from the condition that
$\widehat t_{-1}(u)$ should not have ${\mathbf Q}-$independent term,
$\delta\widehat t_{-1}(u)=0$, leading to
\be
f(u) = \Delta_-(u)\Delta_+(u+i)\,.
\label{f}
\ee
This expression is in agreement with the results of
thorough analysis \cite{MN,Zhou,BPO}.
It is easy to see using \re{fusion} and \re{t-Q} that the
same condition \re{f} ensures that ${\mathbf Q}-$independent
terms do not appear in the expression for the transfer matrix of any
spin.%
\footnote{We would like to stress that this result is
not sensitive to the explicit form of the functions $\Delta_\pm(u)$
and therefore it holds for both open and closed spin chains.}
Substitution of \re{Delta} into \re{f} yields
$$
f(u)=\frac{(2u-i)(2u+3i)}{4u(u+i)}
\delta_+(-u)\delta_-(u) \delta_+(u+i)\delta_-(-u-i)\,.
$$

It is now straightforward to apply the fusion relations,
Eqs.\re{fusion} and \re{f}, to calculate the transfer matrices
of high spins. However, even though one could calculate in the way
$\widehat t_{j}(u)$ for any given fixed $j$, it becomes difficult
to find a general solution to the recurrence relations \re{fusion}
in terms of $\widehat t_{-1/2}(u)$ which in turn are
given by the ABA expression \re{t-M}.
Let us show that this problem can be avoided by using the Baxter
${\mathbf Q}-$function.

\subsection{Eigenvalues of the transfer matrices}

Since the Baxter ${\mathbf Q}-$function satisfies the second-order
finite difference equation \re{Bax}, one expects to find its two
linear independent solutions, ${\mathbf Q}_+(u)$ and ${\mathbf Q}_-(u)$.
One of them, ${\mathbf Q}_+(u)$, provides the polynomial
solution \re{Q-roots} with the roots satisfying the Bethe equations.
The second solution, ${\mathbf Q}_-(u)$, is related to ${\mathbf Q}_+(u)$
through the Wronskian condition
\be
w(u) = {\mathbf Q}_+(u-i) {\mathbf Q}_-(u)
- {\mathbf Q}_+(u) {\mathbf Q}_-(u-i)\,.
\label{W}
\ee
It follows from the Baxter equation \re{Bax} that $w(u)$
satisfies the relations
\be
\frac{w(u+i)}{w(u)}=\frac{\Delta_+(u)}{\Delta_-(u)}\,,\qquad
\frac{w(u+2ij)}{w(u)}=\prod_{k=0}^{2j-1}
\frac{\Delta_+(u+ik)}{\Delta_-(u+ik)}\,.
\label{w-eq}
\ee
As a trivial consequence of the definition \re{W} one
gets the following useful identity
\ba
\frac{{\mathbf Q}_-(u+2ji)}{{\mathbf Q}_+(u+2ji)}
-\frac{{\mathbf Q}_-(u-i)}{{\mathbf Q}_+(u-i)}
=&&
\label{iden}
\\
&& \hspace*{-40mm}
\frac{w(u)}{{\mathbf Q}_+(u-i){\mathbf Q}_+(u)}
+\frac{w(u+i)}{{\mathbf Q}_+(u){\mathbf Q}_+(u+i)}
+ \cdots + \frac{w(u+2ji)}{{\mathbf Q}_+(u+(2j-1)i){\mathbf Q}_+(u+2ji)}\,.
\nonumber
\ea
Applying \re{w-eq} and \re{iden} it is easy to verify that
the spin$-1/2$ and spin$-1$ transfer matrices, Eqs.~\re{t-Q} and \re{t-Q1},
respectively, can be expressed in terms of ${\mathbf Q}_\pm(u)$ as
\ba
&&\widehat t_{-1/2}(u)=\frac{\Delta_-(u)}{w(u)} \left[
{\mathbf Q}_+(u-i){\mathbf Q}_-(u+i)
-{\mathbf Q}_+(u+i){\mathbf Q}_-(u-i)\right]
\\
&&\widehat t_{-1}(u)=\frac{\Delta_-(u)\Delta_-(u+i)}{w(u)} \left[
{\mathbf Q}_+(u-i){\mathbf Q}_-(u+2i)
-{\mathbf Q}_+(u+2i){\mathbf Q}_-(u-i)\right]\,.
\ea
It now becomes straightforward to generalize these expressions to an
arbitrary spin $j$
\be
\widehat t_{-j}(u) = \frac
{\prod_{k=0}^{2j-1} \Delta_-(u+ik)}
{w(u)}
\left[{\mathbf Q}_+(u-i){\mathbf Q}_-(u+2ij)-{\mathbf Q}_+(u+2ij)
{\mathbf Q}_-(u-i)
\right]
\ee
which can be verified to be the exact solution to the fusion
hierarchy \re{fusion}. One can further simplify this expression by using
\re{iden} and writing it entirely in terms of
the polynomial solution ${\mathbf Q}_+(u)\equiv {\mathbf Q}(u)$
\ba
\widehat t_{-j}(u) &=&
{\mathbf Q}(u-i) {\mathbf Q}(u+2ji)
\label{t-j}
\\[3mm]
&\times&
\left[
 \frac{\Delta^{(2j,0)}(u)}{{\mathbf Q}(u-i){\mathbf Q}(u)}
+\frac{\Delta^{(2j-1,1)}(u)}{{\mathbf Q}(u){\mathbf Q}(u+i)}
+ \cdots
+ \frac{\Delta^{(0,2j)}(u)}{{\mathbf Q}(u+(2j-1)i){\mathbf Q}(u+2ji)}
\right]\,.
\nonumber
\ea
Here the notation was introduced for the functions
\ba
\Delta^{(2j,0)}(u)&=&\prod_{k=0}^{2j-1} \Delta_-(u+ki)
\nonumber
\\
\Delta^{(2j-l,l)}(u)&=&\prod_{k=l}^{2j-1} \Delta_-(u+ki)
\prod_{m=0}^{l-1}\Delta_+(u+mi) \,,\quad  1\le l \le 2j-1
\label{Delta's}
\\
\Delta^{(0,2j)}(u)&=&\prod_{m=0}^{2j-1} \Delta_+(u+mi)
\nonumber
\ea
with $\Delta_\pm(u)$ given by \re{Delta}.

Eq.~\re{t-j} is our final expression for the normalized
transfer matrix of the inhomogeneous open spin chain.
One verifies using \re{Q-roots} that for $j=-1/2$ it reproduces
the ABA expression for the spin$-1/2$ transfer matrix \re{t-M}.

\subsection{Eigenvalues of the Hamiltonian}

Calculating the logarithmic derivative of $\widehat t_{-j}(u)$ at $u=-ij$
and putting $j=-j_2$ we can find the eigenvalues of the
Hamiltonian by making use of Eqs.~\re{tj-hat} and \re{H}.
We notice from \re{Delta} and \re{para} that $\Delta_+(u)$ vanishes
at $u=-ij=ij_2$ leading to
\be
\Delta_+(-ij+\epsilon) = {\cal O}(\epsilon^{2(M-2)})\,.
\ee
As a consequence, one finds from \re{Delta's} that
$\Delta^{(2j-l,l)}(-ij+\epsilon)={\cal O}(\epsilon^{2(M-2)})$
for $l=1,...,2j_2$ while
$\Delta^{(2j,0)}(-ij+\epsilon)={\cal O}(\epsilon^0)$
and the expansion of the normalized transfer matrix \re{t-j}
around $u=-ij=ij_2$ looks as
\be
\widehat t_{-j}(-i j + \epsilon) =
\frac{{\mathbf Q}(\epsilon + i j)}{{\mathbf Q}(\epsilon-i j)}
\Delta^{(2j,0)}(-ij+\epsilon) + \CO(\epsilon^{2(M-2)})\,.
\ee
Finally, substituting this expansion into \re{tj-hat} and \re{H}
and putting $j=-j_2$ we calculate the energy of the open spin chain as
\be
{\cal E}_M
=\frac{i}2\left[
\frac{{\mathbf Q}'(i j_2)}{{\mathbf Q}(i j_2)}
-\frac{{\mathbf Q}'(-i j_2)}{{\mathbf Q}(-i j_2)}
\right]+\varepsilon_M\,,
\label{EN1}
\ee
where $j_2=...=j_{M-1}$ and prime denotes a derivative with respect to the
spectral parameter. Here, the constant $\varepsilon_M$ gets contribution from
the $\rho-$factor in \re{tj-hat} as well as the constant term
$\Delta{\cal H}$ in \re{H-open} and determines
the overall normalization of the energy. Since $\varepsilon_M$ does not depend
on the conformal spin of the state $N$, it can be fixed using
the expression for the energy of the pseudovacuum state \re{zero-en}.
The latter is described by a trivial
Baxter ${\mathbf Q}-$function \re{Q-tri}. Substituting
this solution into \re{EN1} we obtain using \re{zero-en}
$$
\varepsilon_M={\cal E}_M(N=0)=0\,.
$$
Taking into account \re{Q-roots} we express the energy \re{EN}
as
\be
{\cal E}_M(N)=i\frac{{\mathbf Q}'(i j_2)}{{\mathbf Q}(i j_2)}
=\sum_{k=1}^N \frac{2j_2}{\lambda_k^2+j_2^2}\,,
\label{EN}
\ee
where we indicated explicitly the dependence of the energy
on the conformal spin $N$ and the number of sites $M$. Remarkably
enough the relation \re{EN}
defining the energy of the {\it open\/} spin chain
coincides with similar expressions for the energy of the {\it closed\/}
spin chain \cite{XXX,FK} although the Bethe roots and the
${\mathbf Q}-$functions are different in two cases.

Eqs.~\re{t-j} and \re{EN} are the main
results of this section. We shall use them in the next section to
find the spectrum of the Hamiltonian ${\cal H}_M$.

\section{Baxter equation}

It follows from our analysis that the Baxter ${\mathbf Q}-$function
determines the spectrum of the transfer matrices,
Eq.~\re{t-j}, and therefore plays a fundamental r\^ole in solving the
eigenproblem for the open spin chains.
Throughout this section we will be mostly interested in solving the
Baxter equation for the spin chain with $M=3$ sites and parameters
defined by the QCD evolution equations \re{j's}. However the methods described
below are general enough and can be applied for solving the Baxter
equation for arbitrary open spin chain.

\subsection{Wilson polynomials}

Before analyzing the Baxter equation for $M=3$ sites let us solve \re{Bax}
in the simplest case of the spin chain with $M=2$ sites. This becomes
useful for a number of reasons. Firstly, the $M=2$ Baxter equation can be
solved exactly. As we will show, its solutions can be identified as Wilson
orthogonal polynomials \cite{AW} and we shall use them as a basis for expanding the
Baxter equation solutions for $M\ge 3$.
Secondly, for the open spin chain with $M=2$ sites the Hamiltonian
\re{H-open} coincides with the pair-wise Hamiltonian \re{H12} and
its diagonalization becomes trivial. As an independent check of
Eq.\re{EN} one should be able to obtain the same
expression for the energy using the exact solution to the $M=2$ Baxter
equation.

To write the Baxter equation \re{Bax} for the spin chain with $M=2$ sites
one needs the expression for the transfer matrix $\widehat t_{-1/2}(u)$ defined
in \re{P3}. Going through the calculation of \re{P3} and \re{abcd}
for $M=2$, we obtain the following relation
\be
\widehat t_{-1/2,M=2}(u)=\Delta_+(u) +\Delta_+(-u)-(4u^2+1) N(N-1+2j_1+2j_2)\,,
\label{t@N=2}
\ee
where $j_1$ and $j_2$ are the spins of the particles, $j_{12}=j_1+j_2+N$
is the conformal spin of the state and $\Delta_+(u)$ is given by
\re{Delta} for $M=2$ as
\be
\Delta_+(u)=\frac{2u+i}{2u}(u-ij_1^+)(u-ij_1^-)(u-ij_2^+)(u-ij_2^-)\,.
\label{Delta-2}
\ee
Eq.\re{t@N=2} allows to rewrite the $M=2$ Baxter equation \re{Bax}
in the form
\be
N(N+2j_1+2j_2-1) {\mathbf Q}(u) =
 \frac{\Delta_+(u)}{4u^2+1}\left[{\mathbf Q}(u-i)-{\mathbf Q}(u)\right]
+\frac{\Delta_+(-u)}{4u^2+1}\left[{\mathbf Q}(u+i)-{\mathbf Q}(u)\right]\,,
\label{Bax-2}
\ee
which has a striking similarity to the definition of the Wilson
polynomials (see Eq.\re{W-eq} in Appendix B). This immediately leads to
\be
{\mathbf Q}_{M=2}(u)
= W_N(u^2;j_1^+,j_1^-,j_2^+,j_2^-)
= W_N(u^2;j_1+\omega_1,j_1-\omega_1,j_2+\omega_2,j_2-\omega_2)\,.
\label{Q=W}
\ee
We conclude that the polynomial solutions to the $M=2$ Baxter equation
\re{Bax} are given by the Wilson orthogonal polynomials with the
parameters fixed by spins and impurities of the model.
This leads to the following property of the Bethe roots, Eq.~\re{Q-roots}.
Being zeros of orthogonal polynomials they are real, simple and interlaced
for different values of the spin $N$. At large $N$ their distribution
on the real axis can be described by a continuous distribution density
\cite{K}.

To obtain the energy ${\cal E}_{M=2}$ we substitute
\re{Q=W} into \re{EN}, identify $\omega=\omega_1-\omega_2$ and put
$\omega_2=0$ for simplicity. Then, using the identity \re{id2} we get
the expression
\baa
{\cal E}_2(N) &=& \psi(j_1+j_2+N+\omega)-\psi(j_1+j_2+\omega)
\\            &+& \psi(j_1+j_2+N-\omega)-\psi(j_1+j_2-\omega)\,,
\eaa
which reproduces the spectrum of two-particle (inhomogeneous) Hamiltonian
\re{H12} for $j_{12}=j_1+j_2+N$.

\subsection{Three-particle Baxter equation}

Let us turn to the Baxter equation for the open spin chain
with $M=3$ sites and choose the spins $j_1=j_3=1$, $j_2=3/2$
to correspond to the QCD evolution equations,
Eqs.~\re{j's}, \re{wS} and \re{wT}. In addition we put $\omega_2=0$
and let the flow parameters $\omega_1$ and $\omega_3$ to vary
according to \re{range}. We obtain from \re{Delta}
\be
\Delta_+(u)=-\frac1{u}\left(u+\frac{i}2\right)
\left(u-\frac32i\right)^2
\prod_{k=1,3}
\left(u-i(1+\omega_k)\right)\left(u-i(1-\omega_k)\right)\,.
\label{Delta-3}
\ee
Evaluating \re{t-hat} one can write the transfer matrix as
\be
\widehat t_{-1/2}(u) = \Delta_+(u) + \Delta_+(-u)+
\left(u^2+\frac14\right) \left[4N(N+6)\left(u^2+\frac34\right)-2q
-15(\omega_1^2+\omega_3^2)+\frac{225}4\right]
\label{t-3}
\ee
where integer $N$ defines the conformal spin of the state \re{h},
and $q=q(\omega_1,\omega_3)$ stands for the eigenvalue of the conserved charge
\re{Q(a)}. These relations allow to represent the Baxter equation
\re{Bax} in the form
\ba
&& \Delta_+(u)\left[{\mathbf Q}(u-i)-{\mathbf Q}(u)\right]
+ \Delta_+(-u)\left[{\mathbf Q}(u+i)-{\mathbf Q}(u)\right]
\label{Bax-3}
\\
&&\hspace*{20mm}
=\left(u^2+\frac14\right) \left[4N(N+6)\left(u^2+\frac34\right)-2q
-15(\omega_1^2+\omega_3^2)+\frac{225}4\right]{\mathbf Q}(u)\,.
\nonumber
\ea
Examining the asymptotics of the both sides of this relation at large
$u$ one finds that ${\mathbf Q}(u) \sim u^{2N}$. Together with
the symmetry of \re{Bax-3} under $u\to -u$ this implies that
${\mathbf Q}(u)$ is given by an even polynomial in $u$ of degree $2N$
\be
{\mathbf Q}(u)= u^{2N} + a_1\, u^{2N-2} + ... + a_N
              = \prod_{k=1}^N (u^2-\lambda_k^2)
\label{Q-gen}
\ee
with $a_k$ being some coefficients and $\lambda_k$ being the Bethe
roots. Substitution of this expansion into
\re{Bax-3} leads to an overcomplete system of linear equations on
the coefficients $a_k$ whose consistency conditions give rise to the
quantization conditions on the charge $q(\omega_1,\omega_3)$.
Their solutions provide the set of quantized $q(\omega_1,\omega_3)$ and
the corresponding Baxter ${\mathbf Q}-$functions. Finally, one
applies \re{EN} to calculate the energy as
\be
{\cal E}_{M=3}(N,q)
=i\frac{{\mathbf Q}'(3i/2)}{{\mathbf Q}(3i/2)}
=\sum_{k=1}^N\frac3{\lambda_k^2+\frac94}
\,.
\label{E-3}
\ee

\subsubsection{Exact solutions}

Although we can not find the general solution to the Baxter equation
\re{Bax-3}, there exist two special values of the flow parameter,
$(\omega_1,\omega_3)=(3/2,1/2)$ and
$(\omega_1,\omega_3)=(3/2,3/2)$, for which an additional
degeneracy occurs and the ${\mathbf Q}-$ function can be calculated exactly.
Remarkably enough it is for these two values of $\omega_1$ and $\omega_3$ that one
recovers the QCD evolution kernels \re{range}. The corresponding
exact anomalous dimensions were first found in \cite{ABH,BBKT} using different
technique.

\bigskip

\noindent
{\it Exact solution for $Q_{S^+}$}.

\bigskip

\noindent
For $(\omega_1,\omega_3)=(3/2,1/2)$
it follows from \re{Delta-3} that $\Delta_+(\pm u)$
acquires the same factor $(u^2+1/4)$ as the r.h.s.\ of \re{Bax-3}.
This allows to simplify \re{Bax-3} as
\ba
&&
\widehat\Delta_+(u)\left[{\mathbf Q}(u-i)-{\mathbf Q}(u)\right]
+\widehat\Delta_+(-u)\left[{\mathbf Q}(u+i)-{\mathbf Q}(u)\right]
\nonumber
\\
&&\hspace*{20mm}
=\left[(4u^2+1)N(N+6)-2(q-q^{(0)}_{S})\right]{\mathbf Q}(u)\,,
\label{Bax-red}
\ea
where
\be
\widehat\Delta_+(u)=-\frac1{u}\left(u+\frac{i}2\right)
\left(u-\frac32i\right)^3\left(u-\frac52i\right)
\ee
and the notation was introduced
\be
q^{(0)}_{S}=N(N+6)+\frac{75}8\,.
\label{QS}
\ee
Putting $u=i/2$ in \re{Bax-red} and using the symmetry property
${\mathbf Q}(-u)={\mathbf Q}(u)$ one finds that the l.h.s.\ of
\re{Bax-red} vanishes leading to
\be
(q-q^{(0)}_{S})\,{\mathbf Q}(i/2)=0\,.
\ee
This relation has two solutions: either $q=q^{(0)}_{S}$,
or ${\mathbf Q}(i/2)=0$.

In the first case, substituting $q=q^{(0)}_{S}$ into \re{Bax-red}
we notice that the ${\mathbf Q}-$function satisfies
the relation that is identical to the Baxter equation \re{Bax-2} for
the spin chain with $M=2$ particles of spins $j_1=3/2$, $j_2=2$
and impurities $\omega_1=0$, $\omega_2=1/2$. This leads to
\be
{\mathbf Q}^{(0)}_{S^+}(u)=
W_N\left(u^2;\frac32,\frac32,\frac32,\frac52\right)\,.
\label{Qs-ex}
\ee
Note that the Bethe roots of \re{Qs-ex} are real. Substituting
${\mathbf Q}^{(0)}_{S^+}(u)$ into \re{EN} and using \re{id2}
we find the energy of the ``exact'' level as
\ba
{\cal E}^{(0)}_{S^+}(N) &=& \psi(N+4)+\psi(N+3)-\psi(4)-\psi(3)
\nonumber
\\
&=&2\psi(N+3)+\frac1{N+3}-\frac{10}3+2\gamma_{_{\rm E}} \,.
\label{ES-ex}
\ea
The explicit expression for the corresponding wave function can be found
in \re{Pol}.

In the second case, for $q\neq q^{(0)}_{S}$ the Baxter
equation \re{Bax-red} can not be solved exactly. Nevertheless the relation
${\mathbf Q}(i/2)=0$ ensures that its possible solutions have a pair
of pure imaginary Bethe roots $\lambda_\pm=\pm i/2$
\be
{\mathbf Q}_{S}(u)=\left(u^2+\frac14\right) P_{N-1}(u^2)
\label{QS-gap}
\ee
with $P_{N-1}(u^2)$ being a polynomial of degree $N-1$ in $u^2$,
while the remaining $2N-2$ roots are real. It follows from \re{E-3}
that the pair of the Bethe roots $\lambda_\pm=\pm i/2$ provides a
contribution to the energy that is independent on the spin $N$ and that
is bigger
than the contribution of a real root $\lambda^2_0>0$ by an amount
\be
\Delta{\cal E}=\frac3{\lambda_\pm^2+9/4}-\frac3{\lambda_0^2+9/4}
> \frac32-\frac43=\frac16\,.
\label{gap-naive}
\ee
Since this contribution is present for all levels except the exact one,
we expect that the energy of the exact level \re{ES-ex}
should be separated from the rest of the spectrum by a finite ``mass gap''.
As we will see in Sect.~7, this is indeed the case and \re{gap-naive} is in
agreement with the large $N$ calculation of the mass gap Eq.~\re{mass-gap}.

\bigskip

\noindent
{\it Exact solution for $Q_{T}$}.

\bigskip

\noindent
For $\omega_1=\omega_3=3/2$ we find from \re{Delta-3} that
$\Delta_+(u)$ vanishes at $u=-i/2$ as $\Delta_+(u)\sim (u+i/2)^3$.
Therefore, expanding the both sides of \re{Bax-3} around $u=-i/2$
up to ${\cal O}((u+i/2)^2)-$order we neglect the terms in \re{Bax-3}
containing $\Delta_+(u)$ and obtain the system of two linear equations
\ba
{\mathbf Q}(i/2)\left[ N(N+6)-q_T-\mbox{$\frac{45}8$}\right] &=&
8i{\mathbf Q}'(i/2)
\label{system}
\\
{\mathbf Q}(i/2) \left[ N(N+6)+q_T+\mbox{$\frac{45}8$}\right] &=&
i{\mathbf Q}'(i/2) \left[ N(N+6)-q_T+\mbox{$\frac{83}8$}\right] \,.
\nonumber
\ea
It has two different solutions: either the charge takes one of the
following two values
\be
q_T^{(+)}= N(N+2)-\frac{45}8\,,\qquad
q_T^{(-)}= N(N+10)+\frac{147}8 \,,
\label{QT}
\ee
or ${\mathbf Q}(i/2)={\mathbf Q}'(i/2)=0$.

In the first case, one is able to find the exact solutions to the
Baxter equation. Similar to \re{Qs-ex}, these solutions are
expressed in terms of the two-particle ${\mathbf Q}-$functions.
However in contrast with \re{Qs-ex}
they are given by the sum of two Wilson polynomials of degree
$N$ and $N-1$, respectively%
\footnote{It is interesting to note that the ${\mathbf Q}_{M=2}-$functions
entering \re{Qs-ex} and \re{QT-ex} are linearly related to each other as
$$
W_N\left(u^2;\mbox{$\frac32,\frac32,\frac32,\frac52$}\right)=
a_{N}
W_N\left(u^2;\mbox{$\frac32,\frac32,\frac52,\frac52$}\right)
+b_N
W_{N-1}\left(u^2;\mbox{$\frac32,\frac32,\frac52,\frac52$}\right)
$$
with $a_N=(N+6)/(2N+6)$ and $b_N=-N(N+2)(N+3)/2$.
}
\be
{\mathbf Q}^{(\pm)}_T(u)=
W_N\left(u^2;\frac32,\frac32,\frac52,\frac52\right)
-\alpha_N^{\pm}
\, W_{N-1}\left(u^2;\frac32,\frac32,\frac52,\frac52\right)\,,
\label{QT-ex}
\ee
where
\be
\alpha^{+}_N=N(N+3)^2\,,\qquad
\alpha^{-}_N=\frac{(N+2)(N+3)^2(N+4)}{N+6}\,.
\ee
We observe that the exact solutions \re{QT-ex} and \re{Qs-ex} are related
to each other as
\be
(N+2)\, {\mathbf Q}_T^{(+)}(u) -N\, {\mathbf Q}_T^{(-)}(u) =
\frac{4(N+3)}{N+6} {\mathbf Q}_{S}^{(0)}(u)\,.
\label{Q-prop}
\ee
Examining the zeros of the polynomials \re{QT-ex} one finds that
${\mathbf Q}_T^{(+)}$ and ${\mathbf Q}_T^{(-)}$ have a pair of pure
imaginary mutually conjugated Bethe roots and the remaining $2(N-1)$
roots are real.

Substituting \re{QT-ex} into \re{EN} and taking into account the
identities \re{id1} and \re{id2} we find the corresponding exact energy
levels as
\ba
{\cal E}_T^{(+)}(N)&=& 2\psi(N+3)-\frac1{N+3}-\frac83+2\gamma_{_{\rm
E}}
\nonumber
\\
{\cal E}_T^{(-)}(N)&=& 2\psi(N+3)+\frac3{N+3}-\frac83+2\gamma_{_{\rm
E}}\,.
\label{ET-ex}
\ea
Note that ${\cal E}_T^{(+)}(0)=0$ in accordance with \re{zero-en}
and the level spacing,
${\cal E}_T^{(+)}(N)-{\cal E}_T^{(-)}(N)=-4/(N+3)$, vanishes at
large $N$.
The property of the ${\mathbf Q}-$functions \re{Q-prop} is translated into
the following relation between the energies
\be
{\cal E}_S^{(0)}(N)=\frac12\left[{\cal E}_T^{(+)}(N)+{\cal
E}_T^{(-)}(N)\right]-\frac23\,,
\label{diff-ex}
\ee
which is valid for arbitrary spin $N$.

As we will see in Sect.~7, the exact solutions, \re{ES-ex} and \re{ET-ex},
are the lowest energy levels in the spectrum of the corresponding
Hamiltonians.
In addition, invariance of the Hamiltonian ${\cal H}_T={\cal H}(3/2,3/2)$
and the charge $Q_T=Q(3/2,3/2)$ under permutations of the end-points
allows to assign a definite parity to the energy levels. One verifies
using the explicit expression for the wave function, Eq.~\re{Pol},
that two exact energy levels, ${\cal E}^{(+)}_T$ and ${\cal
E}^{(-)}_T$ have an opposite parity under $x_1\leftrightarrow x_3$ that
alternates as $N$ changes.

The second solution to the system \re{system}, ${\mathbf Q}(i/2)={\mathbf Q}'(i/2)=0$,
implies that the ${\mathbf Q}-$functions describing ``nonexact''
levels, $q\neq q_T^{(\pm)}$, have a double degenerate pair of pure imaginary
Bethe roots $\pm i/2$
\be
{\mathbf Q}(u) = \left(u^2+\frac14\right)^2 P_{N-2}(u^2)\,.
\ee
Similar to \re{QS-gap}, this property leads to an appearance of the gap
separating the exact levels \re{ET-ex} of the Hamiltonian ${\cal H}_T$ from the
rest of the spectrum.

\subsubsection{Master recurrence relations}

To find the rest of the spectrum of the Hamiltonian
${\cal H}(\omega_1,\omega_3)$ we shall
use the Wilson polynomials as basis for solving the Baxter equation \re{Bax-3}.
Apart from the fact that this basis is very convenient due to
its orthogonality and completeness properties, the expansion of the
$M=3$ Baxter ${\mathbf Q}-$functions over the $M=2$ solutions has a
simple interpretation. It is in one-to-one correspondence with the
decomposition of the 3-body wave function over the 2-particle states.

More explicitly, the wave function of the $M=3$ particle state with given
conformal spin $h=j_1+j_2+j_3+N$ can be decomposed over the set of states
having the same total spin $h$ and, in addition, possessing a definite
conformal spin in the channel defined by any pair of particles.
For instance, the decomposition of the eigenstate $\Psi_N$ in the
$(12)-$channel looks as
\be
\Psi_N(x_i)=\sum_{n=0}^N c_n \Psi_{N,n}^{(12)3}(x_i)\,,
\label{WF-dec}
\ee
where $\Psi_{N,n}^{(12)3}$ has the conformal spin $j_{12}=j_1+j_2+n$
in the channel $(12)$
$$
L_{12}^2 \Psi_{N,n}^{(12)3}(x_i)
= j_{12}(j_{12}-1) \Psi_{N,n}^{(12)3}(x_i)
$$
and $c_n$ are the expansion coefficients. The states $\Psi_{N,n}^{(12)3}$
automatically diagonalize the two-particle Hamiltonian, ${\cal H}_{12}$
and one associates with them the corresponding Wilson polynomials
${\mathbf Q}^{(12)3}_n(u)\equiv W_n(u^2)$. Then, the decomposition of
the $M=3$ solution of the Baxter equation takes the same form as \re{WF-dec}
\be
{\mathbf Q}_N(u) = \sum_{n=0}^N c_n {\mathbf Q}^{(12)3}_n(u)
\label{Q-exp}
\ee
with $c_n$ being the same coefficients as in \re{WF-dec}. The reason for
this is that \re{WF-dec} and \re{Q-exp} describe the {\it same\/}
eigenstate but in different representations. The ${\mathbf Q}-$function
determines the wave function in the separated coordinates
and it is related to \re{WF-dec} through a unitary transformation
well known as the Separation of Variables \cite{SoV}.

Choosing another pair of particles we find different polynomials
${\mathbf Q}^{1(23)}_n(u)$ that
are linear related to ${\mathbf Q}^{(12)3}_n(u)$ through
the Racah $6j-$symbols of the $SL(2)$ group. We obtain their
explicit expressions by replacing
the parameters $j_l$ and $\omega_l$ entering \re{Q=W} by their
values corresponding to the $(12)-$ and $(23)-$pairs of particles
with $j_1=j_3=1$, $j_2=3/2$ and $\omega_2=0$. In this way we get
\ba
{\mathbf Q}^{(12)3}_n(u)&=&
W_n\left(u^2; \frac32,\frac32,1+\omega_1,1-\omega_1\right)
\label{Q12}
\\
{\mathbf Q}^{1(23)}_n(u)&=&
W_n\left(u^2; \frac32,\frac32,1+\omega_3,1-\omega_3\right)\,.
\ea

Substitution of \re{Q-exp} into \re{Bax-3} leads to the system of
recurrence relations on the expansion coefficients $c_k$. To find
their explicit form it proves convenient to define the following
functions
\be
\widetilde\Delta_+(u) = \frac{\Delta_+(u)}{(u-i)^2+\omega_3^2}\,,
\qquad
\widetilde {\mathbf Q}(u) = \left(u^2+\omega_3^2\right){\mathbf
Q}(u)\,.
\label{tilde}
\ee
Here the additional factor was introduced to remove the
contribution to \re{Delta-3} of the particle with the impurity
$\omega_3$. It brings $\widetilde\Delta_+(u)$ to the form
corresponding to the spin chain with $M=2$ sites, Eq.~\re{Delta-2},
and the Baxter equation \re{Bax-3} can be rewritten as
\ba
&&\frac{\widetilde\Delta_+(u)}{4u^2+1}\left[\widetilde {\mathbf Q}(u-i)-
\widetilde {\mathbf Q}(u)\right]
+
\frac{\widetilde\Delta_+(-u)}{4u^2+1}\left[\widetilde {\mathbf Q}(u+i)-
\widetilde {\mathbf Q}(u)\right]
\nonumber
\\
&&\hspace*{40mm}
-(N+1)(N+5)\,\widetilde {\mathbf Q}(u)=
-\frac{q-q_0(N)}{2(u^2+\omega_3^2)}{\mathbf Q}(u)
\label{Q-til}
\ea
with
\be
q_0(N)=N(N+6)\frac{3-4\omega_3^2}{2}-\frac{35}{2}\omega_3^2
-\frac32\omega_1^2+\frac{105}8\,.
\label{Q0(M)}
\ee
Note that the l.h.s.\ of this equation coincides with the two-particle
Baxter equation \re{Bax-2} in the $(12)-$channel and the total two-particle
spin $j_{12}=j_1+j_2+N+1=\frac72+N$. According to \re{Q=W},
the solutions to this reduced equation are given by basis
function ${\mathbf Q}_{N+1}^{(12)3}(u)$ defined in \re{Q12}.
Considering the r.h.s.\ of \re{Q-til} as a perturbation
we seek for a general solution to \re{Q-til} in the form
\be
\widetilde {\mathbf Q}(u) = \sum_{k=0}^{N+1} f_n\,
{\mathbf Q}_n^{(12)3}(u)\,.
\label{Qt-f}
\ee
Substituting this expansion into \re{Q-til} and taking into account the
Baxter equation for ${\mathbf Q}_n^{(12)3}(u)$, Eq.~\re{Bax-2}, we get
\be
2\frac{u^2+\omega_3^2}{q-q_0(N)}\sum_{k=0}^N
(N+1-k)(N+5+k)\, f_k{\mathbf Q}_k^{(12)3}(u)
=\sum_{k=0}^{N+1} f_k {\mathbf Q}_k^{(12)3}(u)\,.
\label{Q-rec}
\ee
Note that according to \re{Qt-f} and \re{tilde} the r.h.s.\ of this relation
is given by $(u^2+\omega_3^2){\mathbf Q}(u)$. This leads to the
following expression for ${\mathbf Q}(u)$  (up to an irrelevant
factor) 
\be
{\mathbf Q}(u)=\sum_{k=0}^N (N+1-k)(N+5+k)\, f_k{\mathbf
Q}_k^{(12)3}(u)\,,
\label{Q-f}
\ee
which coincides with \re{Q-exp} after an appropriate redefinition of
the expansion coefficients.
Applying the identity \re{id3} and using orthogonality of the
Wilson polynomials we find from \re{Q-rec} that the coefficients
$f_k$ satisfy the three-term recurrence relations. To simplify
their form it is convenient to replace $f_k$ by a
new set of coefficients $u_k$ defined as
\be
u_k = f_k (N+1-k)(N+5+k)\, {\mathbf Q}_k^{(12)3}(3i/2)
\ee
with ${\mathbf Q}_k^{(12)3}(3i/2)=
(3)_k\,(\frac52+\omega_1)_k\,(\frac52-\omega_1)_k$
due to
Eqs.\re{Q12} and \re{id1}. Then, the expansion \re{Q-f} takes the form
\be
{\mathbf Q}(u)=\sum_{k=0}^N u_k(q) \frac{{\mathbf Q}_k^{(12)3}(u)}
{(3)_k\,(\frac52+\omega_1)_k\,(\frac52-\omega_1)_k}
=\sum_{k=0}^N u_k(q)\ {}_4F_3\left( {-k,k+4,\frac32+iu,\frac32-iu \atop
3,\frac52+\omega_1,\frac52-\omega_1}\bigg| 1\right)
\label{Q-u}
\ee
with the coefficients $u_k$ satisfying the three-term recurrence relations
\be
-\frac{q-q_0(N)}{2(N+1-k)(N+5+k)} u_k
=
u_{k-1} A_{k-1} - u_k\left(A_k+C_k+\omega_3^2
-\frac94\right)+u_{k+1} C_{k+1}\,.
\label{rec-rel}
\ee
Here $q_0(N)$ was defined in \re{Q0(M)} and the coefficients $A_k$ and $C_k$ are given
by Eqs.\re{A,C} with $a=b=3/2$, $c=1+\omega_1$ and $d=1-\omega_1$
\baa
A_k&=&\frac{(k+3)(k+4)(k+\frac52+\omega_1)(k+\frac52-\omega_1)}{(2k+4)(2k+5)}
\\
C_k&=&\frac{k(k+1)(k+\frac32+\omega_1)(k+\frac32-\omega_1)}{(2k+3)(2k+4)}\,.
\eaa

Substituting \re{Q-u} into \re{E-3} and using the identities \re{id1}
and \re{id2} we find that the expression for the energy takes
a remarkably simple form
\be
{\cal E}(N,q)= \frac{\sum_{k=0}^N u_k(q)
\left[\psi(k+\frac52+\omega_1)+\psi(k+\frac52-\omega_1)
-\psi(\frac52+\omega_1)-\psi(\frac52-\omega_1)
\right]}{\sum_{n=0}^N u_n(q)}\,,
\label{E-rec}
\ee
where we indicated explicitly the dependence of the energy on the
spin $N$ and the conserved charge $q$. This expression has a simple
interpretation. The total energy is given by the sum over all possible
conformal spins $k$ in the two-particle $(12)-$channel. The contribution
of the spin $k$ is proportional to the two-particle energy
${\cal E}_{M=2}(k)$ weighted with the factor $u_k/\sum_n u_n$.

The master recurrence relations \re{rec-rel} allow to construct the solutions
to the Baxter equation \re{Q-u} and, as a consequence, find the spectrum
of the conserved charge $q$ and the energy ${\cal E}(N,q)$.
Solving \re{rec-rel} we impose the normalization condition $u_0(q)=1$.
Under this choice the coefficients $u_n(q)$ are given by polynomials in $q$ of degree
$n$. For $n=N+1$ one finds from \re{rec-rel} that
\be
u_{N+1}(q)=0\,.
\label{u=0}
\ee
This relation provides the quantization
condition on the charge $q$. Solving \re{u=0} we get $N+1$ different
solutions for $q$ that have the properties of roots of the
orthogonal polynomial: quantized $q$ are real, nondegenerate and
interlaced for different values of $N$. Then, it follows from
\re{rec-rel} and \re{E-rec} that the expansion coefficients $u_k(q)$ and the
energy ${\cal E}(N,q)$ are also real.
\begin{figure}
\centerline{\epsfxsize 9cm\epsffile{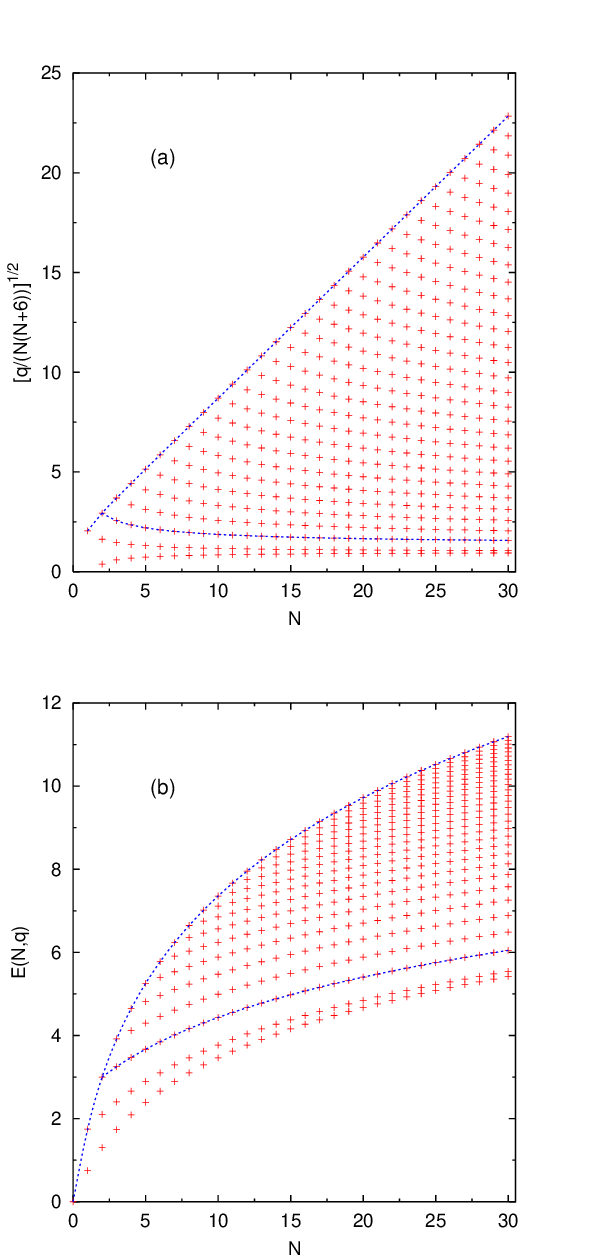}}
\caption[]{\small The dependence of the
quantized values of the charge $q$ and the energy ${\cal E}(N,q)$
on the spin $N$ for the values of the impurity parameters
$\omega_1=\omega_3=3/2$ corresponding to the evolution
kernel ${\cal H}_T$. The dotted lines represent two
trajectories with $\bar\ell=0$ (upper curve) and $\ell=3$ (lower
curve)
described in Sect.~7.4.}\label{Fig:results}
\end{figure}

The results of the numerical solution of the recurrence relations for
the charge $q$ and the energy ${\cal E}(N,q)$ and their dependence on
the spin $N$ and the impurity parameters $\omega_{1,3}$
are shown in Figs.~\ref{Fig:results} and \ref{Fig:flow-en}, respectively.
We observe from Figs.~\ref{Fig:results}
that the eigenvalues of the charge $q$
and the energy ${\cal E}(N,q)$ grow with
the spin $N$ and occupy the band \cite{B}
\be
q_{\rm min}(N) < q < q_{\rm max}(N)\,,
\qquad
{\cal E}_{\rm min}(N) < {\cal E}(N,q) < {\cal E}_{\rm max}(N)
\label{band}
\ee
with $q_{\rm min}={\cal O}(N^2)$, ${\cal E}_{\rm min}\sim 2 \ln N$ and
$q_{\rm max}={\cal O}(N^4)$, ${\cal E}_{\rm max}\sim 4 \ln N$.
In addition, the spectrum of $q$ and ${\cal E}(N,q)$ exhibit an interesting
properties of regularity.
To understand these properties we shall apply in the next section
the asymptotic methods to find an approximate solution to the
Baxter equation for large values of the spin $N$.

\subsubsection{Singular states}

\begin{figure}[t]
\centerline{\epsfxsize 9.2cm\epsffile{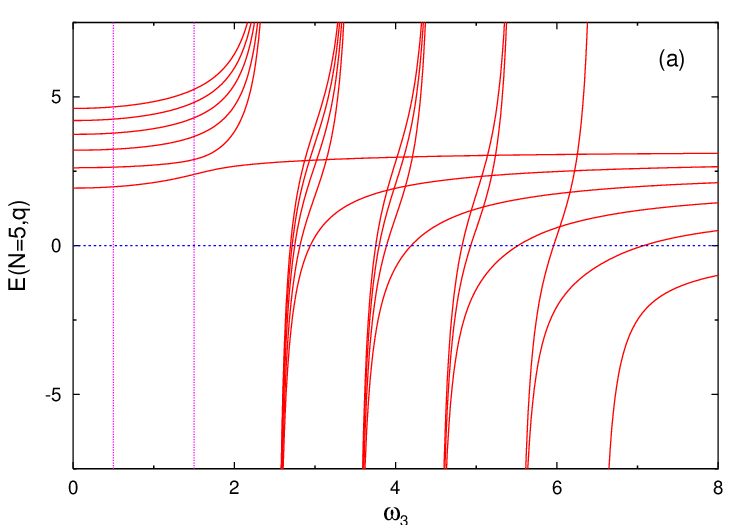}}
\vspace*{3mm}
\centerline{\epsfxsize 9.2cm\epsffile{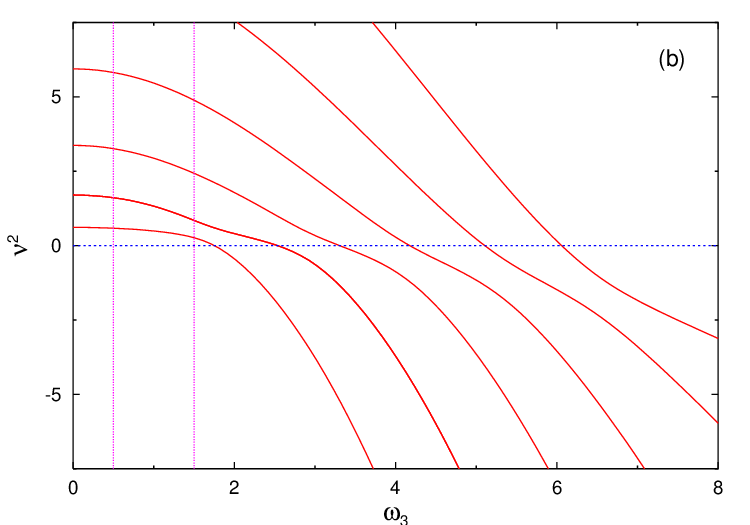}}
\caption[]{\small The flow of the energy ${\cal E}(N,q)$ and the
conserved charge $q=2N(N+6)(\nu^2+3/4)$ with $\omega_3$ for
$N=5$ and $\omega_1=3/2$. Two vertical dotted lines correspond to
$\omega_3=1/2$ and $\omega_3=3/2$ and define the spectrum of the
evolution kernels ${\cal H}_S$ and ${\cal H}_T$, respectively.
}\label{Fig:flow-en}
\end{figure}

According to \re{rec-rel} the solutions to the recurrence
relations depend on the shift parameters $\omega_k$. Examining the
dependence of the energy \re{E-rec} on $\omega_1$ and $\omega_3$
shown in Fig.~\ref{Fig:flow-en}a one observes that ${\cal E}(N,q)$ diverges as
$\omega_k$ approaches half-integer values
\be
\omega_{\rm sing} = \pm\frac52\,,\pm\frac72\,,...\,,\pm\left(\frac32+N\right)\,.
\ee
The total number of the singular points is equal to the spin $N$.
As $N$ increases they start to occupy the whole real axis except of the
interval
\be
-\frac52 < \omega_{1,3} < \frac52\,,
\ee
which defines the stability region of the model.

This phenomenon can be understood by noticing that the two-particle energy in the
$(12)-$channel entering \re{E-rec} diverges as $\omega_k\to\omega_{\rm
sing}$ due to singularities of the $\psi-$functions. Taking
$\omega_1=5/2+n$ with $n=0\,,1\,,...$
we find from \re{E-rec} that divergence comes from
the contribution of two particle spins $k=n+1$, $...$, $N$.
As a consequence, for this value of $\omega_1$ the only levels in the
spectrum that have a finite energy are those with $u_k=0$ for
$k=n+1$, $...$, $N$.
If this condition is not satisfied the energy becomes infinite.%
\footnote{Even though singularities cancel in the numerator of \re{E-rec}
under a weaker condition $\sum_{k=n+1}^N u_k=0$, one finds a stronger
condition $\sum_{k=n+1}^N u_k^2=0$ examining the singular part
of the matrix element of two-particle Hamiltonian in the $(12)-$channel,
$\vev{\Psi_N|H_{12}|\Psi_N}$ with $\Psi_N$ given by \re{WF-dec}.}
For $\omega_1=5/2+n$ the total number of finite energy states
is equal to $n+1$ and at $n=N$ there are no divergent states left in
the spectrum. Since the finite energy states form an orthogonal
subspace, the singular states have $u_k/u_N=0$ for $k=0$, $...$,
$n$.

According to \re{E-3}, the existence of the singular states
corresponds to appearance of the Bethe roots at $\lambda_k=\pm 3i/2$.

\section{Asymptotic solution of the Baxter equation}
\label{Sect:7}

Although the Baxter equation \re{Bax-3} can not be solved exactly
there is a simple way to find its asymptotic solutions at large
values of the spin $N$ \cite{K}. The method is based on the observation
that the both sides of the Baxter equation \re{Bax} have a
different scaling
behaviour at large $N$. We find that for $u={\rm fixed}$ the l.h.s.\
of \re{Bax} scales as $\Delta_\pm(u)\sim N^0$ whereas its r.h.s.\ involves
the transfer matrix \re{t-3} that grows at large $N$ as
\be
\widehat t_{-1/2}(u) = {\cal O}(q)
\label{t-scale}
\ee
depending to the value of the charge $q(N)$, Eq.~\re{band}. Then,
introducing the function
\be
{\bf \varphi}(u)=\frac{{\mathbf Q}(u+i)}{{\mathbf Q}(u)}
\ee
one rewrites the Baxter equation \re{Bax}
\be
\frac{\Delta_+(u)}{\varphi(u-i)} + \Delta_+(-u)\, \varphi(u) =
\widehat t_{-1/2}(u)\,.
\label{phi-eq}
\ee
A general solution to this relation is given by (an infinite)
continuous fraction. Taking into account \re{t-scale} we find
that at large $N$ and $u={\rm fixed}$ the solutions to \re{phi-eq}
are given by
\ba
\varphi_+(u)&=&\frac{\Delta_+(u+i)}{\widehat t_{-1/2}(u+i)}+
{\cal O}(1/q^2)
\,,
\\
\varphi_-(u)&=&\frac{\widehat t_{-1/2}(u)}{\Delta_+(-u)}+
{\cal O}(1/q)
\,.
\ea
It is easy to verify that these relations lead to the following
expressions for the ${\mathbf Q}-$function
\ba
{\mathbf Q}_+(u)&=&2^{i u} \frac{\Gamma(i u)}{\Gamma(i u-\frac12)}
\prod_{k=1}^3 \frac{\Gamma(i(u+\sigma_k))\Gamma(i(u-\sigma_k))}
                   {\Gamma(iu+j_k^+)\Gamma(iu+j_k^-)}
\label{Q+}
\\
{\mathbf Q}_-(u)&=&{\mathbf Q}_+(-u)\,,
\ea
where $j_k^\pm$ were defined in \re{j-pm} and $\sigma_k$ denote
the roots of the
transfer matrix \re{t-3}
\be
\widehat t_{-1/2}(u) = -2 \prod_{k=1}^3 (u^2-\sigma_k^2)\,.
\label{t-roots}
\ee
The general solution to the Baxter equation is given by a
linear combination of ${\mathbf Q}_\pm(u)$. Requiring ${\mathbf Q}(u)$
to be an even function of $u$ one gets
\be
{\mathbf Q}_{\rm as}(u)={\mathbf Q}_+(u)+{\mathbf Q}_-(u)
={\mathbf Q}_+(u)+{\mathbf Q}_+(-u)\,.
\label{Q-as}
\ee
We would like to stress that thus defined asymptotic
${\mathbf Q}-$function obeys the Baxter equation \re{Bax-3} only
in the limit $N\to\infty$ and $u={\rm fixed}$ up to
${\cal O}(1/q)-$corrections.

The dependence of the ${\mathbf Q}-$function on the charge $q$
enters into \re{Q-as} through the roots $\sigma_k$ of the transfer
matrix, Eq.~\re{t-roots}. Since large scales, $q$ and $N(N+6)$, appear in \re{t-3}
with a common prefactor $(u^2+1/4)$, one of the roots is $q-$independent
and it is given by
\be
\sigma_2^2=-\frac14+{\cal O}(N^{-2})\,.
\label{u2}
\ee
Then, taking into account the values of the parameters
$j_k^\pm$ one simplifies \re{Q+} as
\be
{\mathbf Q}_+(u)=2^{iu} \frac{\Gamma(iu)\Gamma(iu+\frac12)}
{\Gamma^2(iu+\frac32)}
\prod_{k=1,3}
\frac{\Gamma(iu-i\sigma_k)\Gamma(iu+i\sigma_k)}
     {\Gamma(iu+1-\omega_k)\Gamma(iu+1+\omega_k)}\,.
\label{Q-sim}
\ee
The values of the remaining roots $\sigma_{1,3}$ vary significantly
as the charge $q$ changes inside the band \re{band}.

Close to the upper bound $q={\cal O}(N^4)$ one finds
that the both roots are real and increase with the spin $N$ as
\be
\sigma_{1,3}^2=\left[1\pm \sqrt{1-\bar q}\right]N(N+6) +{\cal
O}(N^0)\,,
\label{upper}
\ee
where $\bar q=q/(N(N+6))^2={\cal O}(N^0)$.

Close to the lower bound $q={\cal O}(N^2)$ one finds
that one of the roots is $N-$independent
\be
\sigma_1^2=2N(N+6)+{\cal O}(N^0)\,,\qquad
\sigma_3^2=\nu^2+{\cal O}(N^{-2})\,,
\label{lower}
\ee
where the parameter $\nu$ is defined as
\be
\nu^2=\frac{q}{2N(N+6)}-\frac34={\cal O}(N^0)\,.
\label{nu}
\ee
Notice that $\nu$ takes real values for $q \ge 3N(N+6)/2$ and it
becomes pure imaginary for $q < 3N(N+6)/2$. This
happens in particular for the exact levels, Eqs.\re{QS} and \re{QT}
\be
(\nu^{(\pm)}_T)^2=-\frac14\mp\frac2{N} + {\cal O}(N^{-2})\,,\qquad
\nu^2_S=-\frac14+{\cal O}(N^{-2})\,.
\ee
As we will see later, the appearance of the complex roots of the transfer
matrix is closely related to the existence of the mass gap in the spectrum of the
Hamiltonian.

\subsection{Dispersion curve}

Let us apply the solution \re{Q-as} to obtain the asymptotic expression
for the energy ${\cal E}$. According to \re{E-3} the energy is determined
by a logarithmic derivative of the ${\mathbf Q}-$function at $u=\pm 3i/2$.
These values of $u$ belong to the applicability region of the asymptotic
solutions, $N\to\infty$ and $u={\rm fixed}$, and therefore one is allowed
to replace ${\mathbf Q}-$function in \re{E-3} by its asymptotic expression
\re{Q-as}.

It follows from \re{Q-sim} that ${\mathbf Q}_\pm(\pm 3i/2)=0$
and close to $u=\pm 3i/2$ the ${\mathbf Q}-$function is
given by one of the functions, ${\mathbf Q}={\mathbf Q}_\pm(u)$.
Substitution of \re{Q-as} into \re{E-3} yields
\be
{\cal E}_{\rm as}(N,q)=
\sum_{k=1,3}\left[
\psi\left(\mbox{$\frac32$}+i\sigma_k\right)+\psi\left(\mbox{$\frac32$}-i\sigma_k\right)
\right] - \ln 2  - \varepsilon(\omega_1,\omega_3) + {\cal
O}(1/q)\,,
\label{dis}
\ee
where the notation was introduced for the normalization constant
\be
\varepsilon(\omega_1,\omega_3)=\sum_{k=1,3}
  \psi\left(\mbox{$\frac52$}+\omega_k\right)
+ \psi\left(\mbox{$\frac52$}-\omega_k\right)
\label{E0}
\ee
and $\sigma_{1,3}$ are nontrivial roots of the transfer matrix defined
in Eqs.~\re{t-roots} and \re{t-3}. The explicit expressions for
$\sigma_{1,3}$ are quite cumbersome and one can use instead
the approximate expression
\be
\sigma_{1,3}^2 = \left[1\pm \sqrt{1-\widetilde q}\right]N(N+6) +{\cal O}(N^0)
\,,\qquad
\widetilde q= \frac{q-\frac32 N(N+6)}{[N(N+6)]^2}\,,
\ee
which agrees with the asymptotic expressions \re{upper} and
\re{lower}.
The relation \re{dis} establishes the dependence of the energy on the
integrals of motion, $N$ and $q$, and provides an explicit expression
for the dispersion curve of the spin chain model.

The expression \re{dis} simplifies in the upper part of
the spectrum \re{band}. We find using \re{upper} and \re{lower} that
\be
{\cal E}_{\rm as}(N,q)=\ln (q/2) - \varepsilon(\omega_1,\omega_3) + {\cal O}(N^{-2})
\label{E-upper}
\ee
provided that $q={\cal O}(N^4)$.

In the lower part of the spectrum we get from \re{lower}
\be
{\cal E}_{\rm as}(N,q)=\ln[N(N+6)] +
\psi\left(\frac32+i\nu\right)+\psi\left(\frac32-i\nu\right)
-\varepsilon(\omega_1,\omega_3)+{\cal O}(N^{-2})
\label{E-lower}
\ee
with $\nu^2=q/(2N(N+6))-3/4 ={\cal O}(N^0)$. For the special
values of the impurity parameters, Eqs.~\re{wS} and \re{wT},
the relations \re{E-upper} and \re{E-lower} coincide with
analogous expressions obtained in \cite{BDM,B}.

Since $\nu$ could be either real or pure imaginary in \re{E-lower},
we separate the corresponding energy levels into two
groups by introducing the ``vacuum'' energy
\be
{\cal E}_{\rm vac}(N)=\ln[N(N+6)]+2\psi\left(\mbox{$\frac32$}\right)
-\varepsilon(\omega_1,\omega_3)\,.
\label{E-vac}
\ee
The energy levels with $\nu^2\ge 0$ lie above the vacuum,
${\cal E}_{\rm as}(N,q) \ge {\cal E}_{\rm vac}(N)$,
and we shall refer to them as ``continuum'', while a few lowest energy
levels with $\nu^2<0$ and ${\cal E}_{\rm as}(N,q) < {\cal E}_{\rm vac}(N)$
correspond to the ``bound states''.

As we will show in Sect.~7.2.1 (see Eq.~\re{nu-0} below),
the parameters $\nu$ scale in the continuum at large $N$ as $\nu\sim 1/\ln N$.
For small real $\nu$ one expands the r.h.s.\ of \re{E-lower} in powers of
$\nu$ and finds that the energy in the continuum
grows linearly with $q$ close to ${\cal E}_{\rm vac}$
\be
{\cal E}_{\rm continuum}(N,q)-{\cal E}_{\rm vac}(N)=
\left[\frac{q}{N(N+6)}-\frac32\right](8-7\zeta(3))\,.
\ee
The energy of the bound states has a completely different behaviour
at large $N$ because in contrast with the continuum the parameter $|\nu|$
takes {\it finite\/} values for these states. As a consequence,
their energy levels are separated from the continuum by a gap
\be
\Delta {\cal E}={\cal E}_{\rm vac}(N)-{\cal E}_{\rm bound}(N,q)
=2\psi\left(\frac32\right)-
\psi\left(\frac32+|\nu|\right)-\psi\left(\frac32-|\nu|\right)\,.
\ee
We will estimate the value of the mass gap at large $N$ in
Sect.~7.2.3.

\begin{figure}
\centerline{\epsfxsize10.0cm
            \epsffile{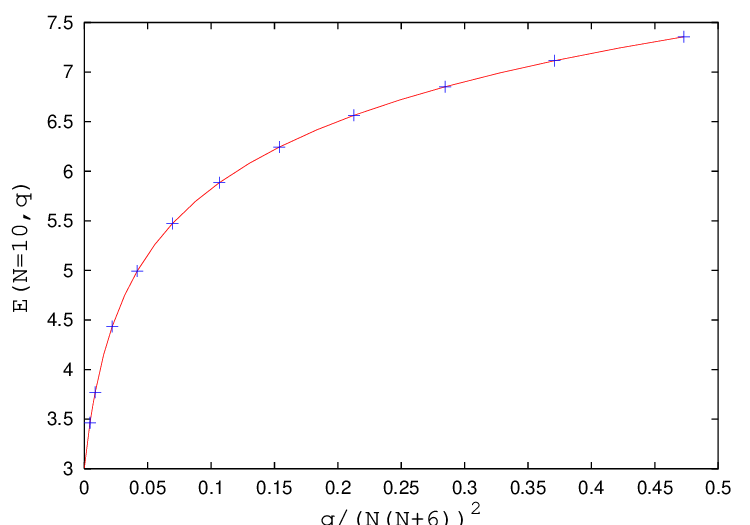}}
\caption[]{\small The dependence on the energy ${\cal E}(N,q)$
on the conserved charge $q$ at $N=10$. Crosses denote the
exact values of the energy and the solid line corresponds to
the asymptotic expression \re{dis}.}
\label{dis-law}
\end{figure}

The asymptotic expansions \re{dis}, \re{E-upper} and
\re{E-lower} are valid up to corrections vanishing
at large $N$. Let us check the accuracy of \re{dis}
by comparing ${\cal E}_{\rm as}(N,q)$ with the exact values
of the energy obtained from the solution of the recurrence
relations, Eqs.~\re{E-rec} and \re{rec-rel}. To this end we
choose the shift parameters to correspond to the QCD evolution kernel
\re{wT}, $\omega_1=\omega_3=3/2$, and take the spin to be $N=10$.
Applying \re{rec-rel} and \re{E-rec} we find $N+1$ pairs
of the exact eigenvalues $(q,{\cal E}_{\rm ex})$ that we
compare with the dispersion curve ${\cal E}_{\rm as}(N,q)$
as shown in Fig.~\ref{dis-law}. Calculating the difference
$\delta{\cal E}={\cal E}_{\rm as}(N,q)-{\cal
E}_{\rm ex}(N,q)$ we find that the asymptotic formula \re{dis}
reproduces the exact energy with a very high accuracy
$\delta{\cal E} < 10^{-5}$, which increases up to
$\delta{\cal E} < 10^{-8}$ as one goes from
the lower part of the spectrum to the upper bound by increasing $q$.

\subsection{Quantization of the conserved charge}

Let us show that the matching of the analytical properties of the
asymptotic solutions \re{Q-as} into Eq.~\re{Q-gen} leads to
the quantization conditions on the charge $q$.

It follows from \re{Q-as} and \re{Q-sim} that the function
${\mathbf Q}_{\rm as}(u)$ has an infinite
series of poles in the complex $u-$plane generated by $\Gamma-$functions
in the numerator of \re{Q-sim} and located at
\be
u_{\rm pole}=\left\{ \pm\frac{i}2\,,\ n \,,\ \pm \sigma_1 + in \,,\
\pm \sigma_3 + in
\right\}
\label{poles}
\ee
with $n$ being arbitrary integer. This seems to be in contradiction with the fact
that the exact solutions to the Baxter equation are given by polynomials
in $u$, Eq.~\re{Q-gen}. Notice however that the asymptotic and the exact
solutions for ${\mathbf Q}(u)$ should coincide (up to corrections
vanishing at large $N$) only in a finite region of $u$ whereas
some poles in \re{poles} are moved outside this region as $N\to\infty$.
This allows to remove from the consideration the ``moving''
poles in \re{poles} satisfying the condition $u_{\rm pole}={\cal O}(N)$ and keep
only the ``fixed'' poles, $u_{\rm pole}={\cal O}(N^0)$. In particular, in the
upper part of the spectrum we get from \re{upper} that
$\sigma_{1,3}={\cal O}(N)$
and therefore the fixed poles of ${\mathbf Q}(u)$ are located at
\be
u_{\rm fixed~ pole}\bigg|_{q={\cal O}(N^4)}=\left\{ \pm\frac{i}2\,,\ in \right\}\,.
\label{poles-up}
\ee
Similarly in the lower part of the spectrum we use \re{lower} and \re{poles}
to find the fixed poles of \re{Q-sim} as
\be
u_{\rm fixed~ pole}\bigg|_{q={\cal O}(N^2)}=\left\{ \pm\frac{i}2\,,\ in \,,
\ \pm \nu+in \right\}
\label{poles-lo}
\ee
with $\nu$ defined in \re{nu} and $n=\rm integer$.

Let us now examine the residue of ${\mathbf Q}_{\rm as}(u)$ at the
fixed poles \re{poles-up} in the upper part of the spectrum.
Using the explicit expression for the functions ${\mathbf Q}_+(u)$,
Eq.~\re{Q-sim}, one finds
that at large $N$ the residue of ${\mathbf Q}_+(u)$ at
$u=im$ with $m=\frac12\,,1\,,2\,,...$ is suppressed with respect to
its residue at $u=0$ by a factor
$(\sigma_1\sigma_3)^{-4m}={\cal O}(q^{-2m})$
that vanishes as $N\to\infty$. Therefore up to ${\cal O}(1/q)-$corrections
one can neglect in \re{poles-up} the fixed poles
at $u=\pm\frac{i}2$ and $u=in$ and keep only the leading pole at $u=0$.

Repeating similar consideration for the fixed poles in the lower
part of the spectrum, \re{poles-lo}, one finds that only the poles
at $u=0$ and $u=\pm \nu$ survive in the large $N$ limit provided
that $|{\rm Im}\,\nu | < 1/2$. Notice that the residue of
${\mathbf Q}(u)$ at $u=\pm \nu$ scales as $\sim q^{\mp 2i\nu}$ and
for $|{\rm Im}\,\nu|\ge 1/2$ it becomes comparable with the contribution
of subleading poles at $u=\pm i/2$ and $u=in\neq 0$.%
\footnote{One can systematically improve the accuracy of the
asymptotic solution by including nonleading $1/q$ corrections to
the asymptotic solutions \re{Q+}.}
Thus, in order to match the asymptotic solution \re{Q-as} into
\re{Q-gen} one has to require that ${\mathbf Q}(u)$ should have
a zero residue at $u=0$. In addition, in the lower part of the
spectrum one has to impose the same condition at $u=\pm i\nu$.

It is easy to see from \re{Q+} that the residue of
${\mathbf Q}_+(u)$ and ${\mathbf Q}_-(u)$ at $u=0$ cancel each
other in the sum \re{Q-as} independently on the value of the
charge $q$. At the same time, calculating the residue of \re{Q-as}
and \re{Q-sim}
at $u=\pm i\nu$ we find after some algebra that it vanishes
provided that $\nu$ obeys the following equation
\be
\left[N(N+6)\right]^{-2i\nu}=
\frac{\Gamma^2(1+2i\nu)\Gamma^2(\frac32-i\nu)}
     {\Gamma^2(1-2i\nu)\Gamma^2(\frac32+i\nu)}
\prod_{k=1,3}
\frac{\Gamma(1+\omega_k-i\nu)\Gamma(1-\omega_k-i\nu)}
     {\Gamma(1+\omega_k+i\nu)\Gamma(1-\omega_k+i\nu)}\,,
\label{Quant-lower}
\ee
where $\nu^2=q/(2N(N+6))-3/4$. This relation establishes the
quantization condition on the charge $q$ in the lower part of
the spectrum $q={\cal O}(N^2)$.

Solving the quantization conditions \re{Quant-lower} we
distinguish two cases: the states in the continuum, $\nu^2>0$,
and the bound states, $\nu^2<0$.

\subsubsection{Solving the quantization conditions in continuum}

For $\nu^2>0$ one gets from \re{Quant-lower}
\be
\nu\ln N(N+6)-{\rm arg}\,
\frac{\prod_{k=1,3}{\Gamma(1+\omega_k+i\nu)\Gamma(1-\omega_k+i\nu)}}
     {\Gamma^2(1+2i\nu)\Gamma^2(\frac32-i\nu)}
= \pi\ell
\label{traj}
\ee
with $\ell$ being an integer. For the special values of the impurity
parameters, Eqs.~\re{wS} and \re{wT}, this relation coincides with the
quantization conditions obtained in \cite{BDM,B}.

It follows from \re{traj} that $\nu$ scales
as $\nu\sim c/\ln[N(N+6)]$ at large $N$ with the constant $c$ having
a nontrivial dependence on the impurity parameters
$\omega_1$ and $\omega_3$. In particular, varying these parameters
one finds that $c$ gets a finite contribution as $\omega_{1,3}$
passes through integer values.
To show this we use the identity
\be
{\rm arg}\, \Gamma(1-\omega_k+i\nu)=
{\rm arg}\, \Gamma(2-\omega_k+i\nu)-
{\rm arg}\, (1-\omega_k+i\nu)
\label{arg1}
\ee
and notice that as $\omega_k$ passes through the value
$\omega_k=1$ the phase of $(1-\omega_k-i\nu)$ changes
by $\pi$. This transition occurs in the region
$|1-\omega_k|={\cal O}(\nu)={\cal O}(1/\ln[N(N+6)])$.
In a similar manner, changing $\omega_k$
from $1$ to $\infty$ we find that every time $\omega_k$ passes integer
positive values the l.h.s.\ of \re{traj} increases by $\pi$, or equivalently
the integer $\ell$ in the r.h.s.\ of \re{traj} decreases as $\ell\to \ell-1$.
As we will see in a moment, this effect corresponds to formation of the
bound state which ``dives'' below the vacuum energy and causes
reparameterization of the levels in the continuum.

For $0\le \omega_1\,,\omega_3 < 1$ and $1-\omega_{1,3}\gg 1/\ln[N(N+6]$
we obtain the solution to \re{traj} as
\be
\nu_0(\ell)=\frac{\pi\ell}{\ln[N(N+6)]+C}
         +{\cal O}\left(\frac{\ell^3}{\ln^3N}\right)
\label{nu-0}
\ee
with
$$
C=4\psi(1)-2\psi(\mbox{$3\over 2$})-\sum_{k=1,3}\left[
\psi(1-\omega_k)+\psi(1+\omega_k)\right]\,.
$$
Substituting \re{nu-0} into \re{E-lower} we find
the energy of the $\ell-$th level as
\be
{\cal E}_\ell = {\cal E}_{\rm vac}
+(14\,\zeta (3)-16)\frac{\pi^2\ell^2}{\ln^2[N(N+6)\,\e^{C}]}
+{\cal O}\left(\frac{\ell^4}{\ln^4 N}\right)
\label{E-l}
\ee
with the vacuum energy ${\cal E}_{\rm vac}$ given by \re{E-vac}.
Integer $\ell=1\,,2\,,...$ enumerates the energy levels with
$\ell=1$ being the lowest level in the continuum. Note that
this expression is valid only for the lowest $l\sim \ln[N(N+6)]$ states.
Using \re{E-l} we find the level spacing in the continuum close to
the vacuum level as
\be
\delta{\cal E}_l(N)\stackrel{\nu\to 0}= {\cal
O}\left(\frac1{\ln^2[N(N+6)]}\right)\,.
\label{cont-space}
\ee

Let us now increase $\omega_1$ and keep $\omega_3$ unchanged.
We find from \re{E-l} that the energy of all levels increases (see
Fig.~\ref{Fig:flow-en}a)
while the difference ${\cal E}_{\ell}-{\cal E}_{\rm vac}$ decreases
with $\omega_1$ (see Fig.\ref{Fig:gap}a).
The lowest level ${\cal E}_{\ell=1}$ rapidly approaches
the vacuum energy ${\cal E}_{\rm vac}$ as $\omega_1\to 1$.
Once $\omega_1$ crosses the value $\omega_1=1$ one applies \re{arg1}
to obtain the solution to \re{traj} as
\be
\nu(\ell)\bigg|_{1 < \omega_1 < 2\atop 0 < \omega_3 < 1} =
\nu_0(\ell-1)
\ee
with $\ell\ge 1$ and $\nu_0(\ell)$ given by
\re{nu-0}. Thus, the lowest energy level $\ell=1$ disappears from the
continuum and this induces reparameterization of the remaining
levels, $\ell\to \ell-1$. Since this effect occurs every time as
$\omega_1$ or $\omega_3$ passes through integer values it is now
easy to write the general solution to \re{traj} valid for arbitrary
values of $\omega_{1,3}$
\be
\nu(\ell)\bigg|_{\omega_1,\,\omega_3 >0 } =
\nu_0(\ell-[\omega_1]-[\omega_3])\,,
\ee
where $[\omega_k]$ denotes an integer part of $\omega_k$.
We conclude that as a result of the flow of the spectrum of the model from
$\omega_1=\omega_3=0$ to arbitrary $\omega_{1,3}$,
\be
{\cal N}_{\rm bound}=[\omega_1]+[\omega_3]
\label{N-bound}
\ee
states cross the vacuum
level with $\nu_{\rm vac}=0$ and the energy ${\cal E}_{\rm vac}$ and
disappear from the continuum. These levels have the charge $\nu^2<0$, or
equivalently $q<3N(N+6)/2$, and they can be described using \re{Quant-lower}.

\subsubsection{Solving the quantization conditions for the bound states}

For $\nu^2 <0$ we put $\nu=i\rho$ with $\rho>0$ and examine the large
$N$ behavior of the both sides of \re{Quant-lower}. The l.h.s.\ of \re{Quant-lower}
grows as
$\sim N^{4\rho}$ whereas the r.h.s.\ is a meromorphic $N-$independent
function of $\rho$. Therefore in the limit $N\to\infty$ one could satisfy
\re{Quant-lower} either through a trivial solution $\rho\sim 1/\ln[N(N+6)]$,
or taking $\rho$ to be close to the poles of the r.h.s.

We find using \re{Quant-lower} that for $\rho>0$ and $0 < \omega_{1,3}<1$ the poles
are located at $\rho=1/2$ and $\rho={\rm positive\ integer}$. However, these
poles coincide with the subleading fixed poles in \re{poles-lo} and their
contribution is beyond an approximation at which \re{Quant-lower} was
obtained.%
\footnote{It follows from the matching condition that subleading corrections
to the asymptotic Baxter equation solution should screen all poles
in \re{poles-lo}.}

Increasing the values of the impurity parameter $\omega_1 >1$ we observe that
the r.h.s.\ of \re{Quant-lower} develops the poles at
\be
\rho=\omega_1-1\,,\omega_1-2\,,...\,, \{\omega_1-1\}
\ee
with $\{\omega_1-1\}$ being a fractional part of $\omega_1$.
Similar phenomenon occurs as one increases the parameter
$\omega_3$. Then, for arbitrary positive $\omega_1$ and $\omega_3$
the total number of the poles is equal to $[\omega_1]+[\omega_3]$
and coincides with the number of the ``missing'' levels from the
continuum ${\cal N}_{\rm bound}$. Thus, at large $N$
Eq.~\re{Quant-lower} has two branches of the solutions each
parameterized by integers $n_1$ and $n_3$, respectively
\be
i\nu = n_k+\{\omega_k-1\}+{\cal O}\left(N^{-1}\right)
\label{nu-k}
\ee
with $k=1,3$ and $n_k=1\,,...\,,[\omega_k-1]$. To find ${\cal O}(1/N)$
corrections to these expressions one has to include subleading corrections to the
quantization condition \re{Quant-lower}, or equivalently to the asymptotic
solution \re{Q-as} and \re{Q+}. Each solution \re{nu-k} corresponds to the bound
state with the energy ${\cal E}_{\rm bound}(n_1)$ or
${\cal E}_{\rm bound}(n_3)$ given by
\be
{\cal E}_{\rm vac}-{\cal E}_{\rm bound}(n_k)
= 2\psi\left(3\over2\right)
- \psi\left({3\over 2}+n_k+\{\omega_k-1\}\right)
- \psi\left({3\over 2}-n_k-\{\omega_k-1\}\right)+{\cal O}\left(N^{-1}\right)
\label{E-bnd}
\ee
with $0\le n_k\le [\omega_k-1]$.
Thus, as $\omega_k$ passes through an integer $n_k$
the level from continuum \re{E-l} with the energy ${\cal E}_{\ell=n_k}$
crosses the vacuum to transform into the bound state with the energy
${\cal E}_{\rm bound}(n_k)$. This transition occurs in the region
$|\omega_k-n_k|={\cal O}(1/\ln N)$ and it is clearly seen on
Fig.~\ref{Fig:gap}a.
For given $\omega_1$ and $\omega_3$ the total number of bound states
is equal to \re{N-bound}. Note that up to ${\cal O}(1/N)$ corrections
the difference ${\cal E}_{\rm vac}-{\cal E}_{\rm bound}$ does not
depend on the spin $N$.

In particular, using \re{N-bound} and \re{nu-k} we find that for
$\omega_1=1/2\,,$ $\omega_3=3/2$
and $\omega_1=\omega_3=3/2$ the spectrum contains ${\cal N}_{\rm bound}=1$
and ${\cal N}_{\rm bound}=2$ bound states, respectively, with
$i\nu=-1/2+{\cal O}(1/N)$. In the second case, two bound states
are degenerate up to ${\cal O}(1/N)$ corrections.
This is in agreement with the exact results \re{ET-ex} and \re{ES-ex}.

\subsubsection{Mass gap}

\begin{figure}[t]
\centerline{\epsfxsize10.0cm
            \epsffile{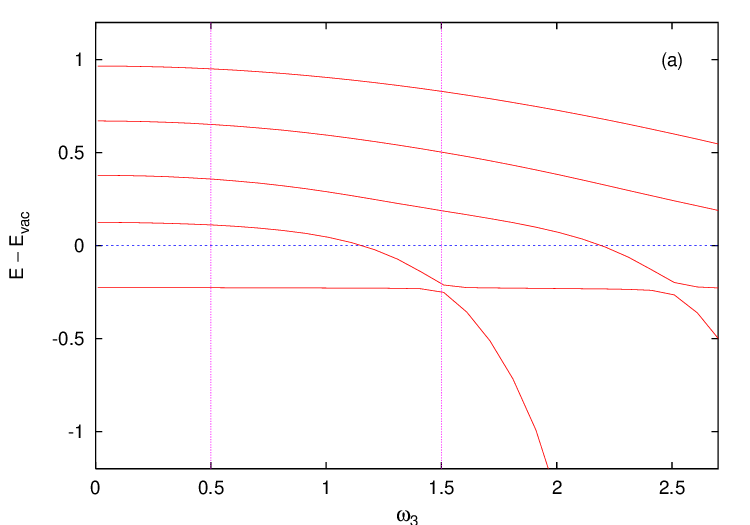}}
\centerline{\epsfxsize10.0cm
            \epsffile{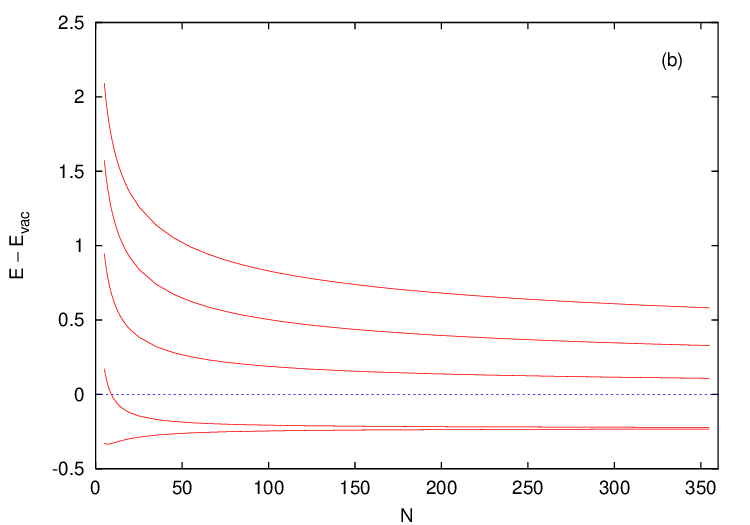}}
\caption[]{\small The difference ${\cal E}(N,q)-{\cal E}_{\rm
vac}(N)$ for the lowest five energy levels of the Hamiltonian
${\cal H}(\omega_1,\omega_3)$: (a) the dependence on the impurity
$\omega_3$ for $N=100$ and $\omega_1=3/2$; (b) the dependence on
the spin $N$ for $\omega_1=\omega_3=3/2$.} \label{Fig:gap}
\end{figure}

We observe from Figs.~\ref{Fig:results}b and \ref{Fig:gap}b that at
large $N$ and fixed $\omega_{1,3}$ the bound states
are separated from the continuum by a finite mass gap,
$\Delta(\omega_1,\omega_3)={\rm min}\left({\cal E}_\ell-{\cal E}_{\rm bound}\right)$.
We estimate its value using \re{E-l} and \re{E-bnd} as
\be
\Delta(\omega_1,\omega_3)={\cal E}_{\rm vac}-{\rm max}_{n_1,n_2}\,{\cal E}_{\rm
bound}(n_k)+{\cal O}(1/\ln^2N)\,.
\ee
Then, $\Delta(\omega_1,\omega_3)=0$ for $0<\omega_{1,3}<1$ and
\be
\Delta(\omega_1,\omega_3)=2\psi\left(3\over2\right)
                         -\psi\left({3\over2}+\{\omega_1-1\}\right)
                         -\psi\left({3\over2}-\{\omega_1-1\}\right)
\label{Delta1}
\ee
both for $0< \omega_3 <1<\omega_1$ and for $1< \omega_1 \le \omega_3$.

Applying \re{Delta1} for the values of parameters $\omega_{1,3}$
corresponding to the QCD evolution kernel, \re{wT} and \re{wS},
one calculates the mass gap as
\be
\Delta\left(\frac32,\frac12\right)=\Delta\left(\frac32,\frac32\right)=
2\psi\left(\mbox{$\frac32$}\right)-\psi(1)-\psi(2)=3-4\ln2
= 0.227411\,.
\label{mass-gap}
\ee
This result agrees with the exact calculations of
the difference ${\cal E}(N,q)-{\cal E}_{\rm vac}(N)$
at $\omega_1=\omega_3=3/2$ as shown in Fig.~\ref{Fig:gap}. Note
that \re{mass-gap} is valid up to corrections
$\sim \pi^2/\ln^2[N(N+6)\e^C]$ which decrease
slowly with $N$ and provide a sizeable contribution to $\Delta$
at finite $N$.

According to \re{mass-gap} the mass gaps in the spectrum of the QCD
evolution kernels, ${\cal H}_T$ and ${\cal H}_{S^+}$, are the same.
Nevertheless, the energies of bound state for these two Hamiltonians
are different due to dependence of the vacuum energy on the shift
parameters, Eqs.~\re{E-vac} and \re{E0}. Their difference is given by
\be
{\cal E}_{T,{\rm bound}}-{\cal E}_{S,{\rm bound}}=
{\cal E}_{\rm vac}\left(\frac32,\frac12\right)-
{\cal E}_{\rm vac}\left(\frac32,\frac32\right)=\frac23
\ee
and it coincides with the exact result \re{diff-ex}.

To summarize, matching the asymptotic solution \re{Q-as}
into the exact form \re{Q-gen} at large $N$ and $u={\rm fixed}$
we obtained the quantization conditions \re{Quant-lower}
and used them to calculate the asymptotic expansion of the charge $q$
and the energy ${\cal E}(N,q)$ in the lower part of the spectrum,
Eqs.~\re{E-l} and \re{E-bnd}, and estimated the value of the
mass gap, Eqs.~\re{Delta1} and \re{mass-gap}.
To find similar expansion in the upper part of the spectrum one needs
an asymptotic solution to the Baxter equation that is valid at large
$N$ and $u={\cal O}(N)$.

\subsection{WKB expansion}

Analyzing the Baxter equation \re{Bax} at large $N$ and
$u={\cal O}(N)$ it is convenient to introduce the scaling
variables
\be
u=\eta x\,,\qquad \eta^2=N(N+6)\,,\qquad \bar q=q/(N(N+6))^2\,.
\ee
Then, one gets from \re{Delta-3} and \re{t-3}
\be
\Delta_+(\eta x) = -\eta^6 x^6\left[1 + i \frac{s}{x}{\eta^{-1}}+{\cal
O}(\eta^{-2})\right]
\ee
with $s=1/2-2\sum_k j_k=-13/2$ and
\be
\widehat t_{-1/2}(\eta x)=2\eta^6 x^6 \left[ V_0(x)-1  +
{\cal O}(\eta^{-2})\right]\,,\qquad V_0=2/x^2-\bar q/ x^4\,.
\ee
Substituting these relations into \re{Bax} one observes
that in the leading $\eta\to\infty$ limit the
Baxter equation takes the form of the discretized one-dimensional
Schr\"odinger equation on the ``wave function'' ${\mathbf Q}(x\eta)$.
In this equation the parameter $1/\eta$ plays the r\^ole of the Planck
constant and $V_0(x)$ defines the potential. This
suggests to look for its solutions in the WKB form \cite{PG}
\be
{\mathbf Q}_+(x\eta)=\exp\left(i\eta S(x)\right)\,,\qquad
S(x)= S_0(x)+\eta^{-1} S_1(x) + {\cal O}(\eta^{-2})\,,
\label{WKB-ans}
\ee
where $S_0(x)$, $S_1(x)$, $...$ are $\eta-$independent
and the expansion is assumed to be uniformly convergent.
Substituting the WKB ansatz \re{WKB-ans} into the Baxter equation \re{Bax}
one expands its both sides in powers of $1/\eta$ to get
\be
\cosh S_0'(x) = 1 - V_0(x)\,,\qquad
S_1'(x)= \frac{i}2 S_0''(x) \coth S_0'(x) -
i\frac{s}{x}\,,\ ...
\label{S_0}
\ee
This leads to the following WKB expansion
\be
{\mathbf Q}_+(x\eta)= \frac{x^{s}}{\sqrt{\sinh
S_0'(x)}}
\exp\left(i\eta \int^x d\xi\, S_0'(\xi)+{\cal O}(\eta^{-1})\right)
\label{Q-WKB}
\ee
with $S_0'(x)$ satisfying \re{S_0}. Finally, combining together
\re{Q-as} and \re{Q-WKB} we arrive at the asymptotic expansion of the
Baxter equation solution that is valid at large $N$ and $u=x\eta={\cal
O}(N)$.

Performing the matching of \re{Q-as} and \re{Q-WKB} into the exact solution
\re{Q-gen}, we shall impose the following two conditions: the
asymptotic solution ${\mathbf Q}_{\rm WKB}(x\eta)$ should be a single
valued function of $x$ and it should oscillate on the real
$x-$axis in order for ${\mathbf Q}-$function to have the real
Bethe roots $\lambda_k={\cal O}(N)$.

Exploring the interpretation of ${\mathbf Q}_{\rm WKB}(x\eta)$ as
a quantum-mechanical WKB wave function we obtain in the standard way that
the first requirement is translated into the Borh-Sommerfeld quantization
condition on the eikonal phase
\be
\frac1{2\pi i}\oint dS(x) = \frac1{\pi} \int_{x_-}^{x_+} dx\, S'(x)
= \bar\ell \eta^{-1}\,,
\label{WKB-q}
\ee
where $\bar l$ is a nonnegative integer and the
integration is performed over a closed contour on the complex $x-$plane
encircling the classical interval of motion $x_-\le x\le x_+$ defined as
\be
\left(1-V_0(x_\pm)\right)^2=
\left(1-2x^{-2}_\pm + \bar q x^{-4}_\pm\right)^2 =1\,.
\label{x-pm}
\ee
Replacing $S=S_0+S_1/\eta$ and using \re{S_0} we get from \re{WKB-q}
\be
\frac1{\pi} \int_{x_-}^{x_+} dx\, S_0'(x) = \left(\bar\ell +
\frac12\right)\eta^{-1} + {\cal O}(\eta^{-2})\,.
\label{WKB-quan}
\ee
Since the WKB wave function oscillates on the interval $[x_-,x_+]$
we can satisfy the second requirement by demanding $x_-$ and $x_+$ to
be real. This leads to the following constraint on the possible values
of the charge $\bar q=q/[N(N+6)]^2$
\be
0 < \bar q \le \frac12\,.
\ee
Moreover, solving the Borh-Sommerfeld quantization conditions
\re{WKB-quan} one can
develop the asymptotic expansion of the charge $q$ close to the upper bound,
$\bar q={\cal O}(N^0)$
\be
\bar q=q/[N(N+6)]^2 = q^{(0)} + \frac{q^{(1)}}{\eta}+
\frac{q^{(2)}}{\eta^2} +\frac{q^{(3)}}{\eta^3}+
\frac{q^{(4)}}{\eta^4} + {\cal O}(\eta^{-5})
\label{Q@WKB}
\ee
with $q^{(k)}$ being the expansion coefficients depending on
integer $\bar\ell$ and the shift parameters $\omega_{1,3}$.
Taking the limit $\eta\to\infty$ in \re{WKB-quan}
one finds that the l.h.s.\ of \re{WKB-quan} has to vanish.
This happens when the classically allowed region shrinks into a
point, $x_+=x_-$, or equivalently
\be
q^{(0)}=\frac12
\label{Q0}
\ee
due to \re{x-pm}.

To calculate the first nonleading coefficient $q^{(1)}$ we have to
find preasymptotic ${\cal O}(\eta^{-1})$ term in the large $\eta-$expansion
of the l.h.s.\ of \re{WKB-quan} and match it against ${\cal O}(\eta^{-1})$ in the r.h.s.\
of \re{WKB-quan}. In this way one gets
\be
q^{(1)}=-\sqrt{2}\left(\bar\ell+\frac12\right)\,.
\label{Q1}
\ee
Note that the coefficients \re{Q0} and \re{Q1} do not depend on the
shift parameters $\omega_{1,3}$.

The WKB expansion \re{Q-WKB} can be systematically improved by taking into account
additional terms in \re{WKB-ans} and using the Baxter equation to express them in terms
of $S_0'(x)$. Going through this procedure and applying the
technique developed in \cite{K} we were able to extend the
expansion \re{Q@WKB} up to ${\cal O}(\eta^{-5})$ term
\baa
q^{(2)}&=&{\frac {155}{16}}+{\frac {11}{8}}\,{\bar\ell}^{2}+{\frac {11}{8}}\,{\bar\ell}
-{\omega_1}^{2}-{\omega_3}^{2}
\\
q^{(3)}&=&\sqrt{2}\left[
-{\frac {47}{128}}\,{\bar\ell}^{3}-{\frac {141}{256}}\,{\bar\ell}^{2}+\left (-{
\frac {3381}{256}}+\frac{\omega_1^2+\omega_3^2}2\right )\bar\ell-{\frac {1667}{
256}}+\frac{\omega_1^2+\omega_3^2}4
\right]
\\
q^{(4)}&=&
{\frac {127}{2048}}\,{\bar\ell}^{4}+{\frac {127}{1024}}\,{\bar\ell}^{3}+\left ({
\frac {3157}{256}}-\frac{\omega_1^2+\omega_3^2}8\right ){\bar\ell}^{2}+\left (
{\frac {25129}{2048}}-\frac{\omega_1^2+\omega_3^2}8\right )\bar\ell
\\
&&+{\frac {94677}{2048}}-{\frac {145}{16}}\,(\omega_1^{2}+\omega_3^2)+\frac{(\omega_1^2-\omega_3^2)^2}2
\,.
\eaa
The asymptotic expansion \re{Q@WKB} describes the spectrum
of the conserved charge $q$ close to the upper bound $q={\cal O}(N^4)$.
It follows from \re{Q@WKB} that quantized values of
the charge form the trajectories $q=q(\bar\ell,N)$ labeled by a
nonnegative integer $\bar\ell$. An example of such a trajectory
for $\bar\ell=0$ is shown in Fig.~\ref{Fig:results}. For the
special values of the impurity parameters, Eqs.~\re{wS} and \re{wT},
the first three terms of the expansion \re{Q@WKB} agree with
analogous expressions obtained in \cite{B}.

For the trajectories with $\bar\ell={\rm fixed}$ the expansion in
\re{Q@WKB} goes over inverse powers of $\eta\sim N$ while for
trajectories with $\bar\ell={\cal O}(N)$ the coefficients $q^{(k)}$
grow as $\sim N^k$. In the latter case, the WKB expansion can
be improved by introducing a new parameter
\be
\xi=\frac{\bar\ell+\frac12}{\eta}\sqrt{2}\,.
\ee
Then, reexpanding $\bar q$ in powers of $1/\eta$ for $\xi={\rm
fixed}$ one finds
\be
\bar q=q_0(\xi)+\eta^{-2} q_1(\xi) + \eta^{-4} q_2(\xi) + {\cal
O}(\eta^{-6})
\label{Wh}
\ee
with
\baa
q_0(\xi)&=&\frac12-\xi+{\frac {11}{16}}\,{\xi}^{2}-{
\frac {47}{256}}\,{\xi}^{3}+{\frac {127}{8192}}\,{\xi}^{4}+{\cal
O}(\xi^5)
\\
q_1(\xi)&=&
{\frac {299}{32}}-{\omega_{{1}}}^{2}-{\omega_{{3}}}^{2}
+\left (\frac{\omega_1^2+\omega_3^2}2-{\frac {6621}{512}}\right )\xi
+\left(-\frac{\omega_1^2+\omega_3^2}{16}+{\frac {50131}{8192}}\right ){\xi}^{2}
+{\cal O}(\xi^3)
\\
q_2(\xi)&=&
{\frac {1414443}{32768}} +\frac{(\omega_1^2-\omega_3^2)^2}2
-{\frac {289}{32}}(\omega_1^2+\omega_3^2) +{\cal O}(\xi)\,.
\eaa
Here, in contrast with \re{Q@WKB} the expansion goes only over
even powers of $\eta$ and it describes the levels with $\bar
l={\cal O}(N)$. Note that the leading term $q_0(\xi)$ does not
depend on the impurity parameters and it can be found exactly as a
solution to the Whitham equations \cite{Wh}.

The obtained WKB expansions \re{Q@WKB} and \re{Wh} combined with the
asymptotic expression for the dispersion curve \re{E-upper} provide a
good description of the spectrum close to the upper bound.

\subsection{Trajectories}

The relations \re{nu-0},\re{nu} and \re{Q@WKB} define two different sets
of the trajectories parameterized by integers $\ell$ and $\bar\ell$,
respectively.
Being combined together they provide a complimentary description
of the spectrum throughout the continuum. Positive integer $\ell$ enumerates
the trajectories \re{E-l} lying above the vacuum level in the lower part
of the spectrum, while integer $\bar\ell=0\,,1\,,2\,,...$ corresponds
to the ordering of the levels from above in the upper part of the spectrum.
Since for fixed spin $N$ the total number of the levels is equal to $N+1$, the
integers $\ell$ and $\bar\ell$ are formally related to each other as
\be
\bar\ell=N-\ell\,.
\ee
The examples of $\bar\ell-$ and $\ell-$trajectories for the
Hamiltonian ${\cal H}_T$ are shown in Fig.~\ref{Fig:results}.
The trajectories with $\bar\ell=0$ and $\ell=3$ go through the
states in the continuum with the maximal amd minimal energy,
respectively. Two trajectories lying below the $\ell=3$ trajectory
correspond to the bound states.

One finds from \re{Q@WKB} and \re{Q1} that the distance between two
neighboring trajectories in the upper part of the spectrum behaves at
large $N$ as
\be
\delta q = [N(N+6)]^2 \delta_{\bar\ell}\,\bar q = {\cal O}(N^3)\,.
\ee
This expression should be compared with the level spacing in the lower
part of the continuum that one finds using \re{nu} and \re{nu-0} as
\be
\delta q =
2N(N+6)\delta_l\,\nu^2={\cal O}\left(\frac{N^2}{\ln^2N}\right)\,.
\ee
Using the properties of the spectral curve, Eq.~\re{E-upper} and \re{E-l},
one obtains the corresponding energy level spacings in the continuum as
\be
\delta {\cal E}(N,q)
\stackrel{q\sim N^2}{=}{\cal O}\left(\frac1{\ln^2 N}\right)\,,\qquad
\delta {\cal E}(N,q) \stackrel{q\sim N^4}{=}{\cal O}\left(\frac1{N}\right)\,.
\label{uni}
\ee
We recall that the bound states are separated
from the continuum by a finite mass gap
\re{Delta1} and \re{mass-gap}.

It is interesting to observe that the same relations \re{uni}
describe the spectrum of the anomalous dimensions of the
baryonic distribution amplitudes \cite{BDKM}. This suggests that
\re{uni} are universal features of the three-particle
evolution equations in multi-color QCD.
We refer to \cite{BDKM} for further
discussion of properties of the trajectories
and their physical interpretation.

Introducing the trajectories one can classify the conformal
operators \re{mult} for different $q$ as belonging to the
different trajectories. Each trajectory describes
a separate component of the twist-3 quark-gluon distribution
\re{mom}. In contrast with the $\mu-$dependence of the distribution,
the scale dependence of its components is of the DGLAP-type,
Eq.~\re{mom}, with the anomalous dimensions given by \re{anom}.
Their mixing with other components is protected by the additional
$Q-$symmetry of the model.

In general, the quark-gluon distributions $D(x_i;\mu)$ enter into
physical observable integrated over parton momentum fractions (see
Eq.~\re{Ys}). As a consequence, it gets contribution from all
trajectories and it scale dependence becomes nontrivial. However
depending on the particular form of the corresponding weight
function one may encounter the situation when most of the
trajectories decouple and only one trajectory contributes in the
multi-color limit. In this case, the scale dependence
significantly simplifies and takes the standard DGLAP form. As
shown in the Appendix A, this is exactly what happens for the
twist-3 nucleon structure functions in the multi-color limit. We
would like to note that this property holds only in the leading
$N_c\to\infty$ limit and the spin structure functions get
contribution from all trajectory through nonplanar
$1/N_c^2-$corrections. These corrections will modify the form of
the twist-3 quark-gluon states constructed in this paper but will
not destroy the analytical properties of the trajectories.

\section{Conclusions}

In this paper we studied the evolution equations for the twist-3
quark-gluon parton distributions in the multi-color QCD.
The evolution equations follow from the scaling
dependence of nonlocal twist-3 quark-gluon string operator on
the light-cone and have the form of one-dimensional
Schr\"odinger equation for three particles on the light-front.
Our analysis was based on the observation that in the multi-color limit
the evolution equations possess an additional integral of motion
and turned out to be effectively equivalent to the Schr\"odinger
equation for integrable open Heisenberg spin chain model. The
parameters of the model are uniquely fixed by properties of
underlying quark-gluon system. Using this correspondence we
constructed the basis of local conformal twist-3 quark-gluon operators
and calculated their anomalous dimensions as the energy levels of the
open spin magnets. We identified the integral
of motion of the spin chain as a new quantum number that separates
different components of the twist-3 parton distributions. Each
component evolves independently and its scaling dependence is
governed by the anomalous dimensions of the conformal operators.

To find the spectrum of the QCD induced open Heisenberg spin
magnet we developed the Bethe Ansatz technique based on the Baxter
equation. Solving a nonlinear fusion relations for the transfer
matrices of the (inhomogeneous) open spin chain models we derived the exact
expression for the energy, or equivalently the anomalous dimensions
of quark-gluon distributions,
in terms of the Baxter ${\mathbf Q}-$function. The properties of the
Baxter equation were studied in detail and its solutions were
constructed using different asymptotic methods. We demonstrated that
the obtained solutions provide a good qualitative description of the
spectrum of the model and reveal a number of interesting
properties: the fine structure of the energy spectrum is described
by the set of trajectories, few lowest energy levels are separated
from the rest of the spectrum by a finite mass gap, for certain
values of the impurity parameters the energy diverges and the system
becomes unstable.

We believe that the open spin chain models define a new universality
class for different problems in high-energy QCD like the Regge asymptotics
of quark-gluon scattering amplitudes and the evolution equations
for high-twist quark-gluon distributions and the results
obtained in this paper could be applied there as well.

Our consideration was restricted to the multi-color limit
$N_c\to\infty$. Nonleading $1/N_c^2-$corrections destroy integrability
of the QCD evolution kernels, Eqs.~\re{ker-T} and \re{ker-S}, and
modify the spectrum of the
anomalous dimensions. Numerical calculations indicate \cite{ABH,B}
that $1/N_c^2$ corrections do not destroy the analytical structure
of the trajectories but modify the level spacing in the upper part
of spectrum generating a mass gap separating the highest energy level
from the rest of the spectrum. These effects can be systematically
taking into account following the approach \cite{BDKM} and deserve
additional studies.

\section*{Acknowledgments}
We are most grateful to V.M.~Braun for illuminating discussions
and for collaboration at the early stage.
One of us (G.K.) would like to thank V.P.~Spiridonov for useful
discussions on the Wilson polynomials.
This work was supported in part by DFG Project N KI-623/1-2 (S.D.)
and by the EU network ``Training and Mobility of Researchers'',
contract FMRX--CT98--0194 (G.K.).

\appendix
\renewcommand{\theequation}{\Alph{section}.\arabic{equation}}

\section{Appendix: Twist-3 nucleon parton distributions}
\setcounter{equation}{0}

In this appendix we summarize the relations between twist-3
nucleon structure functions and quark-gluon distributions
introduced in \re{dis-fun}.

Following \cite{JJ} we define the chiral-odd, $e(x;\mu)$ and
$h_L(x;\mu)$, and the chiral-even, $g_T(x;\mu)$, structure functions
through the matrix elements of nonlocal light-cone quark operators
\ba
&&
\vev{p,s| \bar q(y) q(-y) |p,s} = 2M \int_{-1}^1 dx\,\e^{2ix(py)}
e(x;\mu)
\nonumber\\
&&
\vev{p,s| \bar q(y)y^\mu p^\nu \sigma_{\mu\nu} i\gamma_5 q(-y)|p,s}
=2(sy) M^2\int_{-1}^1 dx\,\e^{2ix(py)}
h_L(x;\mu)
\\
&&
\vev{p,s| \bar q(y) \gamma_\alpha \gamma_5 q(-y)|p,s}
=2Ms_{\perp\alpha}\int_{-1}^1 dx\,\e^{2ix(py)}
g_T(x;\mu)\,,
\nonumber
\ea
where $y_\mu=z n_\mu$ with $n^2=0$ and non-Abelian phase factor is
omitted for brevity. Here, $\ket{p,s}$ is the
nucleon state with momentum $p_\mu$, $p^2=M^2$ and the spin $s_\mu$,
$s^2=-1$. In addition, $s_{\perp\alpha}$ denotes two-dimensional transverse
component of the spin orthogonal to plane defined by the
vectors $n_\mu$ and $p_\mu$.

Using the QCD equations of motion one can separate the twist-2
Wandzura-Wilczek contribution to the structure functions
$h_L(x)$ and $g_T(x)$ \cite{WW,P,SV,CFP,BKL,BB,R}
\ba
h_L(x) &=& 2x\int_{x}^1 \frac{dy}{y^2} h_1(y) + \widetilde h_L(x)
\nonumber
\\
g_T(x) &=& \int_{x}^1 \frac{dy}{y} g_1(y) + \widetilde g_T(x)
\ea
with $h_1(y)$ and $g_1(y)$ being twist-2 nucleon distributions
and $\widetilde h_L(x)$ and $\widetilde g_T(x)$ being genuine
twist-3 part of the structure functions. They can be
expressed as \cite{ABH,BBKT}
\ba
\int_{-1}^1 dx\,\e^{2ix(pn)} e(x;\mu) &=&
\frac{1}{2M}\int_0^1 du \int_{-u}^u dt \vev{p,s|T_{\II}(u,t,-u)|p,s}
\nonumber
\\
\int_{-1}^1 dx\,\e^{2ix(pn)} \widetilde h_L(x;\mu) &=&  -i
\frac{(pz)}{(sz)M^2}\int_0^1 duu \int_{-u}^u dtt \vev{p,s|T_{i\gamma_5}(u,t,-u)|p,s}
\label{str1}
\\
s_{\perp\alpha}\int_{-1}^1 dx\,\e^{2ix(pn)} \widetilde g_T(x;\mu) &=&
-\frac{1}{4M}\int_0^1 du \int_{-u}^u dt
\nonumber
\\
&&
\times \vev{p,s|
(u-t)S_\alpha^+(u,t,-u)+(u+t)S_\alpha^-(u,t,-u)|p,s}\,.
\nonumber
\ea
The matrix element entering the r.h.s.\ of these relations define
(auxiliary) twist-3 quark-gluon distributions \cite{JJ}
\ba
\vev{p,s|T_{\II}(u,t,-u)|p,s} &=&  -4 M (pn)^2\int{\cal D}x\,
\e^{i(pn)[x_1(u-t)-x_3(u+t)]} D_e(x_1,x_2,x_3)
\nonumber
\\
\vev{p,s|T_{i\gamma_5}(u,t,-u)|p,s} &=& 2iM^2(sn)(pn)\int{\cal D}x\,
\e^{i(pn)[x_1(u-t)-x_3(u+t)]} D_{h_L}(x_1,x_2,x_3)
\label{str2}
\\
\vev{p,s|S_\alpha^\pm(u,t,-u)|p,s} &=& 4s_{\perp\alpha} M(pn)^2 \int{\cal D}x\,
\e^{i(pn)[x_1(u-t)-x_3(u+t)]} D_{\pm}(x_1,x_2,x_3)\,.
\nonumber
\ea
Here, ${\cal D}x=dx_1dx_2dx_3\delta(x_1+x_2+x_3)$ with $x_1$, $x_2$
and $(-x_3)$ being the momentum fractions carried by quark, gluon and
antiquark, respectively.

Expanding the both sides of \re{str1} and \re{str2} in powers of $(pn)$
we find the moments of the structure functions as
\ba
\int_{-1}^1 dx x^N e(x) &=& \int_{-1}^1 {\cal D} x D_e(x_i)
\Psi_{N-2}^{(+)}(x_1,x_3)
\nonumber
\\
\int_{-1}^1 dx x^N \widetilde h_L(x) &=& \frac1{N+2}
\int_{-1}^1 {\cal D} x D_h(x_i)
\Psi_{N-2}^{(-)}(x_1,x_3)
\label{str3}
\\
\int_{-1}^1 dx x^N \widetilde g_T(x) &=& \frac1{N+1}
\int_{-1}^1 {\cal D} x
\left[D_+(x_i)\Psi_{N-2}^{(0)}(x_1,x_3)
-D_-(x_i)\Psi_{N-2}^{(0)}(x_3,x_1)\right]\,,
\nonumber
\ea
where $\Psi_{N}^{(\pm,0)}(x_i)$ are homogeneous polynomials
in $x_1$ and $x_3$ of degree $N$ defined as
\ba
\Psi_N^{(+)}(x_1,x_3) &=& \frac{x_1^{N+1}-(-x_3)^{N+1}}{x_1+x_3}
\nonumber
\\
\Psi_N^{(-)}(x_1,x_3) &=& \left[\partial_{x_1}+\partial_{x_3}\right]
\frac{x_1^{N+2}-(-x_3)^{N+2}}{x_1+x_3}
\label{Pol}
\\
\Psi_N^{(0)}(x_1,x_3) &=& \partial_{x_1}\frac{x_1^{N+2}-(-x_3)^{N+2}}
{x_1+x_3}\,.
\nonumber
\ea
Inverting these relations and taking into account the spectral
properties of the distributions one gets \cite{CFP,BKL,JJ,R}
\ba
e(x)&=&\frac1{x}\int_{-1}^1 \frac{dx'}{x-x'}Y_e(x,x')
\nonumber
\\
\widetilde h_L(x) &=& -x\int_{x}^1 \frac{dx'}{x'^2}
\int_{-1}^1 \frac{dx''}{x'-x''}
\left[\partial_{x'}-\partial_{x''}\right]
Y_h(x',x'')
\label{Ys}
\\
\widetilde g_T(x)&=&
-\int_{x}^1 \frac{dx'}{x'}
\int_{-1}^1 \frac{dx''}{x'-x''}
\left[\partial_{x'}Y_g(x',x'')+\partial_{x''}Y_g(x'',x')\right]
\,,
\nonumber
\ea
where the notation was introduced for (anti)symmetrized twist-3 quark-gluon
distributions
\baa
Y_{e}(x',x'')&=&D_{e}(x',-x'+x'',-x'')+ D_{e}(x'',x'-x'',-x')
\\
Y_{h}(x',x'')&=&D_{h}(x',-x'+x'',-x'')- D_{h}(x'',x'-x'',-x')
\\
Y_g(x',x'')&=&D_+(x',-x'+x'',-x'')+D_-(x'',x'-x'',-x')\,.
\eaa
In contrast with the $D_{e,h,g}-$functions these distributions are
real functions of the parton fractions.

One can verify by a direct calculation that the polynomials \re{Pol}
diagonalize the conserved charges $Q_T$ and $Q_{S^+}$, Eqs.~\re{int-T} and
\re{int-S}, on the subspace $x_1+x_2+x_3=0$
\be
Q_T \Psi_N^{(\pm)}(x_i)=q_T^{(\pm)} \Psi_N^{(\pm)}(x_i)\,,
\qquad
Q_{S^+} \Psi_N^{(0)}(x_i)=q_S^{(0)} \Psi_N^{(0)}(x_i)
\ee
with the eigenvalues $q_T^{(\pm)}$ and $q_S^{(0)}$ given by
the exact solutions, Eqs.~\re{QT} and \re{QS}, respectively. Then,
comparing \re{str3} with \re{mom} we conclude that in the multi-color
limit, $N_c\to\infty$, the moments of the
nucleon structure functions,
${\cal M}_f(N;\mu)=\int_{-1}^1 dx x^N f(x;\mu)$,
 have a scale dependence of
the DGLAP type
\be
\mu\frac{d}{d\mu} {\cal M}_f(N;\mu)= 
-\gamma_f(N){\cal M}_f(N;\mu) +{\cal O}(1/N_c^2)\,,
\qquad f=e\,,\widetilde h_L\,,\widetilde g_T
\ee
with the anomalous dimensions equal to the energies of
the exact levels, Eqs.~\re{const}, \re{ES-ex} and \re{ET-ex}.
For the chiral-odd structure functions one gets \cite{BBKT}
\ba
\gamma_{e}(N)&=&\frac{\alpha_s N_c}{2\pi}
\left[ {\cal E}_{T}^{(+)}(N-2) +\frac{13}6\right]+{\cal O}(1/N_c^2)
\nonumber
\\
\gamma_{\widetilde h_L}(N) &=& \frac{\alpha_s N_c}{2\pi}
\left[{\cal E}_{T}^{(-)}(N-2)+\frac{13}6\right]+{\cal O}(1/N_c^2)
\ea
and for chiral-even structure function  one
finds the flavor nonsinglet (NS) contribution to the anomalous
dimension as \cite{ABH}
\be
\gamma_{\widetilde g_T}^{\rm NS}(N)=
\frac{\alpha_s N_c}{2\pi}
\left[{\cal E}_{S}^{(0)}(N-2)+\frac{17}6\right]+{\cal O}(1/N_c^2)
\,.
\ee
This simplification occurs because among all possible components of
the twist-3 quark-gluon distributions the twist-3 nucleon structure
functions ``select'' only those corresponding to the lowest exact
levels. The
scale dependence of the chiral-odd structure functions $e(x)$ and
$\widetilde h_L(x)$ is associated with two ``exact'' trajectories of
the Hamiltonian ${\cal H}_T$, while the chiral-even structure function
$\widetilde g_T(x)$ is governed by the ``exact'' trajectory of the
Hamiltonian ${\cal H}_S$.

\section{Appendix: Wilson polynomials}
\setcounter{equation}{0}

In this Appendix we summarize the properties of the Wilson
polynomials \cite{AW}. They are defined in terms of hypergeometric series
as
\be
W_n(x^2;a,b,c,d)=(a+b)_n (a+c)_n (a+d)_n\ {}_4F_3\left(
{{-n,n+s-1,a+ix,a-ix} \atop {a+b,a+c,a+d}} \bigg| 1 \right)
\label{Wil}
\ee
with $s=a+b+c+d$, $(a)_n\equiv \Gamma(a+n)/\Gamma(a)$, $n$ integer and
$a$...$d$  being arbitrary parameters. $W_n$ are given
by polynomials in $x^2$ of degree $n$ and the prefactor
in \re{Wil} is chosen to ensure the symmetry with
respect to any permutations of the parameters
\be
W_n(x^2;a,b,c,d)= W_n(x^2;b,a,c,d)= ... = W_n(x^2;a,b,d,c)\,.
\ee
The Wilson polynomials form the system orthogonal polynomials
on a line $-\infty < x< \infty$ and the corresponding orthogonality and
completeness conditions can be found in \cite{AW}.
They satisfy the second-order finite difference equation
\be
n(n+s-1) y(x) = B(x) \left[y(x+i)-y(x)\right]
              + B(-x)\left[y(x-i)-y(x)\right]
\label{W-eq}
\ee
with $y(x)=W_n(x^2;a,b,c,d)$ and
$$
B(x)=-\frac{(x+ia)(x+ib)(x+ic)(x+id)}{2x(2x+i)}\,.
$$
In addition, $W_n$ obeys the following three term recurrence relations
\be
x^2 P_n(x^2) = P_{n+1}(x^2)+(A_n+C_n-a^2) P_n(x^2)
             + A_{n-1} C_n P_{n-1}(x^2)\,.
\label{id3}
\ee
Here,
$$
P_n(x^2) = (-1)^n\frac{(s-1)_n}{(s-1)_{2n}} W_n(x^2;a,b,c,d) = x^{2n} + ...
$$
are normalized polynomials and
\ba
A_n&=&\frac{(n+s-1)(n+a+b)(n+a+c)(n+a+d)}{(2n+s-1)(2n+s)}
\nonumber
\\
C_n&=&\frac{n(n+b+c-1)(n+b+d-1)(n+c+d-1)}{(2n+s-2)(2n+s-1)}\,.
\label{A,C}
\ea
Using \re{Wil} one can obtain the following useful relations
\ba
W_n(x^2;a,b,c,d)\bigg|_{x=\pm ia} &=& (a+b)_n (a+c)_n (a+d)_n
\label{id1}
\\
\pm i\partial_x \ln W_n(x^2;a,a,c,d)\bigg|_{x=\pm ia} &=&
\psi(n+a+c)-\psi(a+c)
\label{id2}
\\
&+&\psi(n+a+d)-\psi(a+d)\,.
\nonumber
\ea


\end{document}